\documentclass[trackchanges, twocolumn]{aastex7}

\usepackage{comment}
\usepackage{subcaption}
\usepackage{graphicx}
\usepackage{array}
\newcommand{\notecell}[1]{\parbox[t]{0.64\textwidth}{\raggedright\footnotesize #1}}
\usepackage{amsmath}
\usepackage{multirow}

\begin{document}

\title{The JADES Transient Survey III: Linking Core-Collapse Supernova Rates to Cosmic Star Formation}

\author[0000-0001-5517-6335]{Christian Vassallo}
\affiliation{Tuorla Observatory, Department of Physics and Astronomy, University of Turku, 20014 Turku, Finland}
\email{clvass@utu.fi}

\author[0000-0001-7497-2994]{Seppo Mattila}
\affiliation{Tuorla Observatory, Department of Physics and Astronomy, University of Turku, 20014 Turku, Finland}
\affiliation{School of Sciences, European University Cyprus, Diogenes Street, Engomi, 1516 Nicosia, Cyprus}
\email{sepmat@utu.fi}

\author[0000-0002-4781-9078]{Christa DeCoursey}
\affiliation{Steward Observatory, University of Arizona, 933 N. Cherry Avenue, Tucson, AZ 85721, USA}
\email{cndecoursey@arizona.edu}

\author[0000-0002-7756-4440]{Louis-Gregory~Strolger}
\affiliation{Space Telescope Science Institute, 3700 San Martin Drive, Baltimore, MD 21218, USA}
\email{strolger@stsci.edu}

\author[0000-0001-8257-3512]{Erkki Kankare}
\affiliation{Tuorla Observatory, Department of Physics and Astronomy, University of Turku, 20014 Turku, Finland}
\email{erkki.kankare@utu.fi}

\author[0000-0002-6842-3021]{Max M. Briel}
\affiliation{Département d’Astronomie, Université de Genève, Chemin Pegasi 51, CH-1290 Versoix, Switzerland}
\affiliation{Gravitational Wave Science Center (GWSC), Université de Genève, CH-1211 Geneva, Switzerland}
\email{max.briel@gmail.com}

\author[0000-0003-1344-9475]{Eiichi Egami}
\affiliation{Steward Observatory, University of Arizona, 933 N. Cherry Avenue, Tucson, AZ 85721, USA}
\email{egami@arizona.edu}

\author[0009-0003-2907-1873]{Iikka Mäntynen}
\affiliation{Tuorla Observatory, Department of Physics and Astronomy, University of Turku, 20014 Turku, Finland}
\email{Iikka.mantynen@windowslive.com}

\author[0000-0003-4263-2228]{David~A.~Coulter}
\affiliation{Department of Physics and Astronomy, The Johns Hopkins University, 3400 N. Charles St., Baltimore, MD 21218, USA}
\affiliation{Space Telescope Science Institute, 3700 San Martin Drive, Baltimore, MD 21218, USA}
\email{dcoulter@stsci.edu}

\author[0000-0002-4410-5387]{Armin~Rest}
\affiliation{Space Telescope Science Institute, 3700 San Martin Drive, Baltimore, MD 21218, USA}
\email{arest@stsci.edu}

\author[0000-0002-8651-9879]{Andrew J.\ Bunker}
\affiliation{Department of Physics, University of Oxford, Denys Wilkinson Building, Keble Road, Oxford OX1 3RH, U.K.}
\email{andy.bunker@physics.ox.ac.uk}

\author[0000-0002-0450-7306]{Alex J.\ Cameron}
\affiliation{Cosmic Dawn Center (DAWN), Copenhagen, Denmark}
\affiliation{Niels Bohr Institute, University of Copenhagen, Jagtvej 128, DK-2200 Copenhagen, Denmark}
\email{alex.cameron@nbi.ku.dk}

\author[0000-0002-2929-3121]{Daniel J.\ Eisenstein}
\affiliation{Center for Astrophysics $|$ Harvard \& Smithsonian, 60 Garden St., Cambridge, MA 02138, USA}
\email{deisenstein@cfa.harvard.edu}

\author[0000-0003-2238-1572]{Ori D.\ Fox}
\affiliation{Space Telescope Science Institute, 3700 San Martin Drive, Baltimore, MD 21218, USA}
\email{ofox@stsci.edu}

\author[0000-0003-4565-8239]{Kevin Hainline}
\affiliation{Steward Observatory, University of Arizona, 933 N. Cherry Avenue, Tucson, AZ 85721, USA}
\email{kevinhainline@arizona.edu}

\author[0000-0002-8543-761X]{Ryan Hausen}
\affiliation{Department of Physics and Astronomy, The Johns Hopkins University, 3400 N. Charles St., Baltimore, MD 21218, USA}
\email{rhausen@ucsc.edu}

\author[0000-0001-7673-2257]{Zhiyuan Ji}
\affiliation{Steward Observatory, University of Arizona, 933 N. Cherry Avenue, Tucson, AZ 85721, USA}
\email{zhiyuanji@arizona.edu}

\author[0000-0002-9280-7594]{Benjamin D.\ Johnson}
\affiliation{Center for Astrophysics $|$ Harvard \& Smithsonian, 60 Garden St., Cambridge, MA 02138, USA}
\email{benjamin.johnson@cfa.harvard.edu}

\author[0000-0002-4985-3819]{Roberto Maiolino}
\affiliation{Kavli Institute for Cosmology, University of Cambridge, Madingley Road, Cambridge, CB3 0HA, UK}
\affiliation{Cavendish Laboratory -- Astrophysics Group, University of Cambridge, 19 JJ Thomson Avenue, Cambridge, CB3 0HE, UK}
\affiliation{Department of Physics and Astronomy, University College London, Gower Street, London WC1E 6BT, UK}
\email{rm665@cam.ac.uk}

\author[0000-0003-1169-1954]{Takashi~J.\ Moriya}
\affiliation{National Astronomical Observatory of Japan, National Institutes of Natural Sciences, 2-21-1 Osawa, Mitaka, Tokyo 181-8588, Japan}
\affiliation{Graduate Institute for Advanced Studies, SOKENDAI, 2-21-1 Osawa, Mitaka, Tokyo 181-8588, Japan}
\affiliation{School of Physics and Astronomy, Monash University, Clayton, VIC 3800, Australia}
\email{takashi.moriya@nao.ac.jp}

\author[0000-0002-2361-7201]{Justin~D.~R.~Pierel}
\affiliation{Space Telescope Science Institute, 3700 San Martin Drive, Baltimore, MD 21218, USA}
\email{jpierel@stsci.edu}

\author[0000-0002-1022-6463]{Thomas M. Reynolds}
\affiliation{Tuorla Observatory, Department of Physics and Astronomy, University of Turku, 20014 Turku, Finland}
\affiliation{Cosmic Dawn Center (DAWN), Copenhagen, Denmark}
\affiliation{Niels Bohr Institute, University of Copenhagen, Jagtvej 128, DK-2200 Copenhagen, Denmark}
\email{thomas.reynolds@nbi.ku.dk}

\author[0000-0002-4271-0364]{Brant Robertson}
\affiliation{Department of Astronomy and Astrophysics, University of California, Santa Cruz, 1156 High Street, Santa Cruz, CA 95064, USA}
\email{brant@ucsc.edu}

\author[0000-0002-4622-6617]{Fengwu Sun}
\affiliation{Center for Astrophysics $|$ Harvard \& Smithsonian, 60 Garden St., Cambridge, MA 02138, USA}
\email{fengwu.sun@cfa.harvard.edu}

\author[0000-0002-8224-4505]{Sandro Tacchella}
\affiliation{Kavli Institute for Cosmology, University of Cambridge, Madingley Road, Cambridge, CB3 0HA, UK}
\affiliation{Cavendish Laboratory -- Astrophysics Group, University of Cambridge, 19 JJ Thomson Avenue, Cambridge, CB3 0HE, UK}
\email{st578@cam.ac.uk}

\author[0000-0003-2919-7495]{Christina C.\ Williams}
\affiliation{NSF National Optical-Infrared Astronomy Research Laboratory, 950 North Cherry Avenue, Tucson, AZ 85719, USA}
\email{christina.williams@noirlab.edu}

\author[0000-0001-9262-9997]{Christopher N.\ A.\ Willmer}
\affiliation{Steward Observatory, University of Arizona, 933 N. Cherry Avenue, Tucson, AZ 85721, USA}
\email{cnaw@as.arizona.edu}

\begin{abstract}
We investigate how core-collapse supernova (CCSN) rates trace the star-formation rate densities (SFRDs) over the redshift range $0 \le z \le 5$. For this we use new high-redshift results from the James Webb Space Telescope Advanced Deep Extragalactic Survey (JADES) Transient Survey (JTS, see companion paper by DeCoursey et al. 2026), together with published CCSN rates. Using the observed CCSN rates to constrain the CCSN production efficiency relating SFRDs to CCSN rates, we examine how the inferred connection between star formation rates and CCSN production efficiency depends on the stellar initial mass function (IMF) and the adopted CCSN progenitor mass range. We find that the observed CCSN rates are consistent with dust extinction-corrected UV+IR–based SFRDs for plausible CCSN progenitor masses. Using the observed CCSN rates to directly reconstruct the cosmic star-formation history, we recover a peak at $z_{\rm peak}=2.3^{+0.6}_{-0.5}$, in agreement with galaxy luminosity-based determinations. Allowing the IMF to evolve with redshift has only a modest impact when SFRD estimates are treated consistently, indicating that CCSN rates are not as sensitive to the change of IMF as might be assumed. Adopting higher SFRDs that include a dust-obscured population of faint millimeter sources implies a substantial and increasing fraction of missing, dust-obscured CCSNe at higher redshifts. Although the inferred fraction of CCSNe missed by the surveys depends on the adopted CCSN production efficiency, we find an increasing fraction of supernovae missed due to obscuration, rising from modest values at low redshift to a peak at z $\sim2$, and remaining substantial toward z $\sim5$.
\end{abstract}


\section{Introduction}

Star formation plays a central role in galaxy evolution, governing the build-up of stellar mass, the chemical enrichment of the interstellar medium, and the energetic feedback processes that regulate subsequent galaxy evolution (e.g. \citet{Madau2014}). Quantifying how the star-formation rate (SFR) evolves with cosmic time is therefore a fundamental goal of observational cosmology. Recent advances in observational capabilities across a broad range of wavelengths, enabled by both ground-based and space-based facilities, have substantially improved our understanding of the origin and evolution of galaxies (e.g., \citealt{robertson}).

The cosmic SFR density (SFRD) is most commonly inferred from rest-frame
ultraviolet (UV) emission tracing young, massive stars, recombination
lines such as H$\alpha$, and infrared (IR) emission from dust that
re-radiates absorbed starlight. These tracers are typically combined and
corrected for dust attenuation to estimate the total SFRD
(e.g. \citealt{Kennicc, Madau2014}). While such measurements have led to a
well-established picture in which cosmic star formation peaks at
$z \simeq 1.9$, uncertainties remain due to calibration systematics,
assumptions about dust attenuation, and the potential incompleteness of
even UV+IR-based measurements.

Core-collapse supernovae (CCSNe) are directly connected to ongoing star formation since they result from short-lived massive stars when the nuclear fuel has been exhausted. The lifetime of such massive stars from their formation to core collapse is $\lesssim 50$ Myr \citep{Zapartas17, Briel2022}, which is negligible compared to the Gyr timescales, over which the cosmic star formation rates evolve \citep{Madau2014}. Therefore, by investigating the CCSN rates, a relation can be established between the rate of CCSNe and the underlying star formation rates \citep[e.g.][]{DahlenFransson, cappellaro1999}. This provides an additional method to probe SFRs independent of galaxy luminosities and to study the populations of massive stars leading to CCSNe as a function of redshift.

However, the relation between the CCSN rate and the SFRD is not straightforward. The conversion factor of the CCSN--SFRD relation depends on the assumed initial mass function (IMF) and on the mass range of stars that produce CCSNe. Uncertainties in the lower and upper progenitor mass limits propagate directly into the CCSN production efficiency, and therefore into the expected CCSN rate per unit SFR \citep[e.g.][]{Botticella2012}. The lower ZAMS mass limit for single star CCSN progenitors is generally assumed to be around $\sim 8 \pm1 \mathrm{M_\odot}$ \citep{Smartt}. Direct detections of CCSN progenitors have suggested an upper limit of progenitor ZAMS masses around $\sim$18–20 $\mathrm{M_\odot}$ \citep{Smartt_2015}, but systematic effects and binary stellar evolution effects may shift the inferred upper limit to $\sim$25–30 $\mathrm{M_\odot}$ or higher \citep[e.g.,][]{Beasor2016,eldridge2013,vandyk2017}. In addition, both the SFRD and CCSN rate measurements are affected by dust obscuration. While galaxy luminosity-based SFRs are typically corrected for dust attenuation using UV slopes and infrared (IR) measurements, these corrections may not fully capture the most heavily obscured star formation. Similarly, CCSN rate studies commonly apply statistical corrections for host-galaxy extinction but may still miss CCSNe that are intrinsically faint or occur in very dusty environments.

Comparisons between CCSN rates and cosmic SFRD have previously revealed a potential discrepancy, with observed CCSN rates appearing lower than expected (\citealt{Horiuchi2011a}). This tension has been attributed to CCSNe missed by rest-frame optical surveys, particularly in dusty starbursts \citep[e.g.,][]{VanBuren1994,mattila2001} and luminous and ultraluminous infrared galaxies (LIRGs and ULIRGs; \citealt{Perez}), where a substantial fraction of CCSNe may be obscured (e.g., \citealt{Mannucci2007, Mattila2012a, miluzio2013, Kool2018, Kankare2021, Fox2021, mantynen2025}). Several studies have shown that after correcting for this obscured CCSN population, the inferred CCSN rates are consistent with expectations from the cosmic SFRD (e.g., \citealt{Dahlen2012a, Melinder2012, cappellaro2015}).

At the same time, recent ALMA-based studies suggest that even
dust extinction-corrected UV+IR measurements may underestimate the true cosmic SFR
density, particularly at high redshift
(e.g. \citealt{Fujimoto24}). If both star formation and CCSNe are
increasingly hidden by dust at early cosmic times, then both SFRD and
CCSN rates may be systematically underestimated, complicating direct
comparisons between these two.

Recently, JWST has been shown to have the ability to discover the highest redshift SNe \citep[][]{decoursey,yan25,fox26}. In particular, \citet{decoursey} report that the JWST Advanced Deep Extragalactic Survey (JADES; \citealt{eisenstein, rieke}) has uncovered some of the highest redshift spectroscopically confirmed SNe (Types Ia, Ic-BL and II-P; e.g., \citealt{pierel, siebert, coulter, moriya}) known so far, with the most distant event reported by \citet{Coulter2026} from the cluster-lensing VENUS survey.

This work builds on the volumetric CCSN rates derived from JADES observations and presented in a companion paper (DeCoursey et al. submitted), and focuses on interpreting these rates in the context of cosmic star formation history and dust obscuration. The newly derived CCSN rates from the JWST JADES Transient Survey (JTS), combined with CCSN rate measurements from the
literature, are used to investigate the relation between CCSN rates and SFRDs from the local Universe to $z \sim 5$. We address three closely related questions: (i) whether dust extinction-corrected SFRDs reproduce the observed CCSN rates for reasonable assumptions about the CCSN progenitors; (ii) what cosmic star formation history is implied when observed CCSN rates are used as SFRD tracers, and (iii) what redshift dependent fraction of CCSNe must have been missed due to dust obscuration when adopting the higher ALMA-based SFRDs.

This paper is organized as follows. In Section \ref{sec:observedrate}, we describe the CCSN rate data used in this work. In Section \ref{sec:LinkingSFR}, we compare SFR-derived CCSN rates with observations and the effects of the IMF. In Section \ref{sec:constrains}, we infer the cosmic star-formation history directly from the observed CCSN rates. In Section \ref{sec:MF}, we constrain the redshift-dependent missing CCSN fraction. In Section \ref{sec:systematics} we discuss JTS sample selection effects and the effect Hubble constant on the expected rates and summarize our findings in Section \ref{sec:conclusions}.

\section{Core-Collapse Supernova Rates in the Literature}
\label{sec:observedrate}
The analysis presented in this work is based on a compilation of volumetric CCSN rate measurements spanning from the local Universe to $z \sim 5$. This compilation combines newly derived high-redshift CCSN rates from the JWST JADES Transient Survey (JTS) (C. DeCoursey et al. submitted) with published CCSN rates from the literature at lower redshifts. Together, these measurements provide a continuous observational baseline for studying the relation between CCSN rates and the cosmic star-formation history over most of cosmic time.

The JTS enables the detection of CCSNe at redshifts
previously inaccessible to optical surveys, owing to the depth and
near- to mid-infrared wavelength coverage of the JWST. The JTS CCSN rates extend to $z \sim 5$ and anchor the high-redshift end of the CCSN rate evolution used throughout this paper. At lower redshifts, we include CCSN rate measurements from a range of ground- and space-based surveys. These surveys differ in wavelength coverage, cadence, and detection strategy, but collectively trace the evolution of the CCSN rate over $0 \lesssim z \lesssim 2$. 

The full CCSN rate compilation used in this work is listed in
Table~\ref{tab:ccsn_rates} and presented in Figure \ref{fig:CCSNe_andk}. Our main analyses use the full JTS sample, with the "Gold" sample results discussed in section 6.1. The rates are reported in commonly used standard units and, unless otherwise noted, assume the same cosmological parameters ($H_0 = 70~\mathrm{km\,s^{-1}\,Mpc^{-1}}$, $\Omega_M = 0.3$, and $\Omega_\Lambda = 0.7$), with the exception of \citet{Strolger2015a}, who adopt $\Omega_M = 0.27$ and $\Omega_\Lambda = 0.73$. We describe the previous highest redshift CCSN rate results below. For a full description of the JTS sample and the inferred rates, see \cite{decoursey} and C. DeCoursey et al. submitted.

\cite{Dahlen2012a} presented a sample of 45 CCSNe discovered with the Advanced Camera for Surveys (ACS) on HST to estimate the CCSN rates over $0.1<z<1.3$. Two CCSNe were classified spectroscopically and the remainder photometrically using the $F606W$, $F775W$, and $F850LP$ filters. The rates were derived in three different redshift bins centered at $z=0.39$, 0.73, and 1.11, and were reported with and without corrections for host galaxy extinction and CCSNe missed in dusty U/LIRGs. We adopt the host galaxy corrected rates to allow for a customized redshift-dependent missing-SN correction. The authors showed that systematic uncertainties increase significantly beyond $z>0.5$, driven primarily by dust extinction and the SN luminosity function, with systematic errors dominating over statistical uncertainties at higher redshifts, but remaining negligible in their lowest redshift bin.

We also adopt the host galaxy extinction corrected CCSN rates from \citet{Melinder2012}, who provided the rates over $0.1<z<1.0$ based on $R$ and $I$ band deep imaging using the European Southern Observatory (ESO) Very Large Telescope (VLT) as a part of the Stockholm VIMOS Supernova Survey (SVISS). These rates were based on the detection of a total of nine CCSNe classified based on their light curve and colors.

We adopt CCSN rates from \cite{Petrushevska2016}, who measured CCSN rates at $0.4 \lesssim z \lesssim 2.9$ using a near-infrared search behind the galaxy cluster Abell~1689, taking advantage of gravitational lensing to probe higher redshifts. Their results were based on multi-epoch $J$-band observations with the HAWK-I instrument on the VLT, with supporting optical data from the Nordic Optical Telescope. Five photometrically classified CCSNe were detected at $0.67 < z < 1.70$, and volumetric CCSN rates were derived including host-galaxy extinction corrections. The inferred rates are consistent, within large uncertainties, with expectations based on the cosmic star-formation history. Since the highest-redshift bin is reported as an upper limit, we exclude this point and include only the host extinction-corrected measurements with positive detections in our analysis.

\cite{Strolger2015a} presented CCSN rates using data from the Cosmic Assembly Near-infrared Deep Extragalactic Legacy Survey (CANDELS \citealt{Grogin, koekemoer}) and the Cluster Lensing And Supernova survey with Hubble (CLASH \citealt{Postman}), observed with ACS onboard the HST. The filters used were F125W and F160W (similar to the $J$ and $H$ bands, respectively). The study extended the measurements of the CCSN rates to a redshift of 2.5, based on a sample of approximately 44 mostly photometrically classified CCSNe. 
Systematic uncertainties arose from various factors, including fractions of different SN types, mean absolute peak magnitudes, and adopted k-corrections. 
For example, variations in relative rates between different SN types can reduce the volumetric rates by up to 20\%, while differences in absolute peak magnitudes can cause discrepancies as large as 45\% in the highest redshift bins. We use the statistical uncertainties reported in their Table 4 and the systematic uncertainties from their Table 3 (normal galaxies) by \cite{Strolger2015a}.

\begin{table}[!htbp]
\centering
\caption{Compilation of CCSN rates used in this work}
\begin{tabular}{cccc}
\hline
\hline
Redshift & Rate* & Stat. $\pm$ (Sys.)** & Reference \\
\hline
0.0025 & 1.50 & $^{+0.40}_{-0.30}\,(0.0)$ & Mattila12 \\
0.003  & 1.63 & $^{+0.41}_{-0.41}\,(0.0)$ & Botticella12 \\
0.005  & 0.69 & $^{+0.08}_{-0.08}\,(^{+0.21}_{-0.03})$ & Ma25 \\
0.0087 & 0.62 & $^{+0.07}_{-0.07}\,(^{+0.13}_{-0.11})$ & Li11 \\
0.010  & 0.43 & $^{+0.17}_{-0.17}\,(0.0)$ & Cappellaro99$^\dagger$ \\
0.015  & 0.70 & $^{+0.10}_{-0.09}\,(0.0)$ & Pessi25$^\dagger$ \\
0.030  & 1.01 & $^{+0.50}_{-0.30}\,(0.0)$ & Perley20$^\dagger$ \\
0.030  & 0.91 & $^{+0.16}_{-0.13}\,(0.0)$ & Frohmaier21$^\dagger$ \\
0.072  & 1.06 & $^{+0.11}_{-0.11}\,(\pm0.15)$ & Taylor14 \\
0.075  & 1.04 & $^{+0.33}_{-0.26}\,(^{+0.04}_{-0.11})$ & Graur15 \\
0.10   & 1.13 & $^{+0.62}_{-0.53}\,(\pm0.49)$ & Cappellaro15 \\
0.21   & 1.15 & $^{+0.43}_{-0.33}\,(^{+0.42}_{-0.36})$ & Botticella08 \\
0.25   & 1.21 & $^{+0.27}_{-0.27}\,(\pm0.47)$ & Cappellaro15 \\
0.26   & 2.20 & $^{+0.80}_{-0.70}\,(0.0)$ & Cappellaro05$^\dagger$ \\
0.30   & 2.13 & $^{+0.80}_{-0.57}\,(^{+0.98}_{-0.78})$ & Strolger15 \\
0.39   & 2.46 & $^{+2.30}_{-1.33}\,(^{+1.75}_{-1.42})$ & Melinder12 \\
0.39   & 2.24 & $^{+0.96}_{-0.70}\,(^{+0.46}_{-0.37})$ & Dahlen12 \\
0.57   & 2.8 & $^{+4.5}_{-2.0}\,(^{+0.8}_{-0.8})$ & Petrushevska16 \\
0.66   & 6.90 & $^{+2.54}_{-2.78}\,(^{+7.38}_{-4.60})$ & Graur11 \\
0.70   & 3.86 & $^{+0.96}_{-0.72}\,(^{+1.26}_{-0.53})$ & Strolger15 \\
0.73   & 4.21 & $^{+3.49}_{-2.05}\,(^{+2.94}_{-1.98})$ & Melinder12 \\
0.73   & 4.86 & $^{+1.22}_{-1.00}\,(^{+1.36}_{-0.93})$ & Dahlen12 \\
1.08   & 10.3 & $^{+11.5}_{-6.1}\,(^{+3.3}_{-3.0})$ & Petrushevska16 \\
1.10   & 3.07 & $^{+1.06}_{-0.66}\,(^{+0.83}_{-0.74})$ & Strolger15 \\
1.10   & 5.95 & $^{+2.34}_{-1.74}\,(^{+1.81}_{-1.60})$ & Dahlen12 \\
1.50   & 3.25 & $^{+2.03}_{-1.32}\,(^{+1.62}_{-1.69})$ & Strolger15 \\
1.57   & 10.8 & $^{+26.4}_{-9.6}\,(^{+3.8}_{-3.7})$ & Petrushevska16 \\
1.90   & 3.16 & $^{+3.37}_{-1.77}\,(^{+1.65}_{-1.55})$ & Strolger15 \\
2.30   & 6.17 & $^{+6.76}_{-3.52}\,(^{+4.67}_{-3.94})$ & Strolger15 \\
\hline
\multicolumn{4}{l}{JADES (JTS) CCSN rates} \\
\hline
  $0.95$--$1.49$ & 4.5 & $^{+1.8}_{-1.3}\,(^{+0.3}_{-0.7})$ & JTS \\
  $1.49$--$2.05$ & 5.6 & $^{+2.1}_{-1.6}\,(^{+0.4}_{-0.5})$ & JTS \\
  $2.05$--$2.78$ & 6.5 & $^{+2.3}_{-1.7}\,(^{+0.5}_{-0.3})$ & JTS \\
  $2.78$--$5.20$ & 4.1 & $^{+1.4}_{-1.1}\,(^{+1.0}_{-0.3})$ & JTS \\
\hline
\end{tabular}

\vspace{1ex}
{\footnotesize
* Rates are in units of $10^{-4}h_{70}^{3}\mathrm{yr}^{-1}\mathrm{Mpc}^{-3}$, where the volume is comoving. \\
** Uncertainties are split into statistical and systematic, the latter are shown in parentheses. \\
$\dagger$ Measurements not corrected for host-galaxy dust extinction and therefore treated as lower limits (see Table~\ref{tab:host_extinction}). \\[0.5ex]

References:  
\cite{Mattila2012a};
\cite{Botticella2008};
\cite{Botticella2012};
\cite{Botticella2017};
\cite{Ma2025};
\cite{li2011};
\cite{cappellaro1999};
\cite{Cappellaro2005};
\cite{cappellaro2015};
\cite{Graur2011};
\cite{Graur2015};
\cite{Dahlen2012a};
\cite{Strolger2015a};
\cite{Melinder2012};
\cite{Petrushevska2016};
\cite{perley2020};
\cite{Frohmaier2021a};
\cite{Pessi2025};
JTS: \cite{decoursey};
DeCoursey et al. 2026, submitted.
}

\label{tab:ccsn_rates}
\end{table}

The published CCSN rates also differ in their treatment of host-galaxy extinction and dust-obscured events. Some studies use host-galaxy extinction corrections, while others report observed optical rates that should be interpreted as lower limits. We summarize the treatment of host-galaxy extinction by each survey in Appendix Table \ref{tab:host_extinction}.

\begin{table}[t]
\centering
\caption{Cosmic star-formation rate density parameterizations used in this work.}
\label{tab:sfrd_params}
\begin{tabular}{lcccc}
\hline
\hline
Ref. & $A$ & $B$ & $C$ & $D$ \\
\hline
MD14     & 0.015 & 2.7 & 2.9 & 5.6 \\
F24 & $0.010 \pm 0.003$ & $2.8 \pm 0.5$ & $3.3 \pm 0.3$ & $6.6 \pm 0.5$ \\
\hline
\end{tabular}
\vspace{1ex}
{\footnotesize
MD14: \cite{Madau2014} \\
F24: \cite{Fujimoto24} \\
SFRD parameterization of the form in Equation \ref{eq:sfr_params}.}
\end{table}

\section{Linking Star Formation Rates to CCSN Rates}
\label{sec:LinkingSFR}

To compare CCSN rates with expectations from SFRDs, we adopted parameterized descriptions of the CSFH from the literature. In particular, we consider the widely used compilation of \citet{Madau2014}, which combines rest-frame UV and IR measurements and applies dust attenuation corrections to infer the total SFRD as a function of redshift. These SFRDs are commonly interpreted as dust-corrected estimates of the total star-formation activity.

We also consider the UV-based measurements of \citet{Harikane22}, who derived rest-frame UV luminosity functions and clustering constraints from $\sim4\times10^6$ galaxies at $z\sim2$–7 using Subaru/HSC data. The \citet{Harikane22} CSFH broadly follows the shape of \citet{Madau2014}, but with updated luminosity functions and improved high-redshift statistics.

Finally, we include the ALMA-based SFRD measurements of \citet{Fujimoto24}, who studied 180 dust-continuum sources detected in 33 massive galaxy clusters as part of the ALCS survey, covering 133 arcmin$^2$ at 1.2\,mm. Gravitational lensing enables the detection of faint millimeter sources, including near-IR-dark galaxies with no HST or \textit{Spitzer} counterparts. Their total SFRD at $z>4$ is estimated to be $\sim1.6$ times higher than UV-based, extinction-corrected measurements, highlighting a significant dust-obscured component. They fit the functional form of Equation~\ref{eq:sfr_params} \citep{Madau2014} to their data, anchoring the low-redshift normalization with the $z=0$ point from \citet{Madau2014}.

The SFRD parameterizations adopted in this work are summarized in Table~\ref{tab:sfrd_params}. The \citet{Harikane22} SFRD is not included in this table, as their CSFH (cosmic star formation history) parametrization follows a different functional form (see their Equation~60) than the MD14-style formulation adopted here. These models are used in the following sections to predict CCSN rates and to compare with the CCSN-derived star-formation histories.

\subsection{Expected CCSN Rates from SFRDs}
\label{sec:LinkingSFR2}

Given an SFRD, the expected volumetric
CCSN rate can be calculated by assuming an IMF and a range of stellar masses for stars that explode as CCSNe.
Because the lifetimes of massive stars are short compared to timescales at which the cosmic SFRD evolves, the CCSN rate at a given redshift is directly proportional to the SFR in the same redshift range.

The expected CCSN rate is given by
\begin{equation}
\label{eq:ccsn_sfr_relation}
\begin{aligned}
R_{\mathrm{CCSN}}(z)
&=
\frac{\int_{m_l}^{m_u} \phi(m)\,\mathrm{d}m}
     {\int_{m_{\min}}^{m_{\max}} m\,\phi(m)\,\mathrm{d}m}
\,\rho_{\mathrm{SFR}}(z) \\
&\equiv
k_{\mathrm{CCSN}}\,\rho_{\mathrm{SFR}}(z)\, .
\end{aligned}
\end{equation}

where $\phi(m)$ is the IMF, $m_l$ and $m_u$ define the CCSN progenitor ZAMS mass range, and $m_{\min}$ and $m_{\max}$ define the full stellar mass range of the IMF considered. Here, $\rho_{\mathrm{SFR}}(z)$ denotes the SFRD as a function of redshift, expressed in units of $\mathrm{M_\odot} \mathrm{yr}^{-1} \mathrm{Mpc}^{-3}$. The proportionality constant $k_{\mathrm{CCSN}}$ represents the number of CCSNe produced per unit stellar mass formed and has units of $\mathrm{M_\odot^{-1}}$. In this formulation we make a simplifying assumption that the CCSN outcome is assumed to be determined by a single ZAMS mass cut, effectively treating stars above a chosen mass threshold as failing to produce an observable CCSN. We do not include the non-monotonic islands of explodability seen in some simulations (e.g., \citealt{OConnor2011, Sukhbold2016, Patton2020}). We explore this effect in C. Vassallo et al. in (preparation). In this work, we also assume no redshift dependence for $k_{\mathrm{CCSN}}$.

In this section, we adopt parameterized SFRDs from the literature and use Equation~\ref{eq:ccsn_sfr_relation} to predict the expected CCSN rate as a function of redshift. We also repeat this analysis using CCSN rates corrected for supernovae missing in heavily dust obscured galaxies, applying the prescription of \citet{Mattila2012a}, extrapolated beyond $z=2$ under the assumption that the correction remains constant. We derive an independent estimate of the missing supernova fraction by comparing the observed CCSN rates to the ALMA-based SFRD of \citet{Fujimoto24}, as described in Section \ref{sec:MF}

\subsection{Best-fit $k_{\mathrm{CCSN}}$ Values}
\label{sec:LinkingSFR3}

\begin{figure*}
    \centering
    \includegraphics[width=0.9\linewidth]{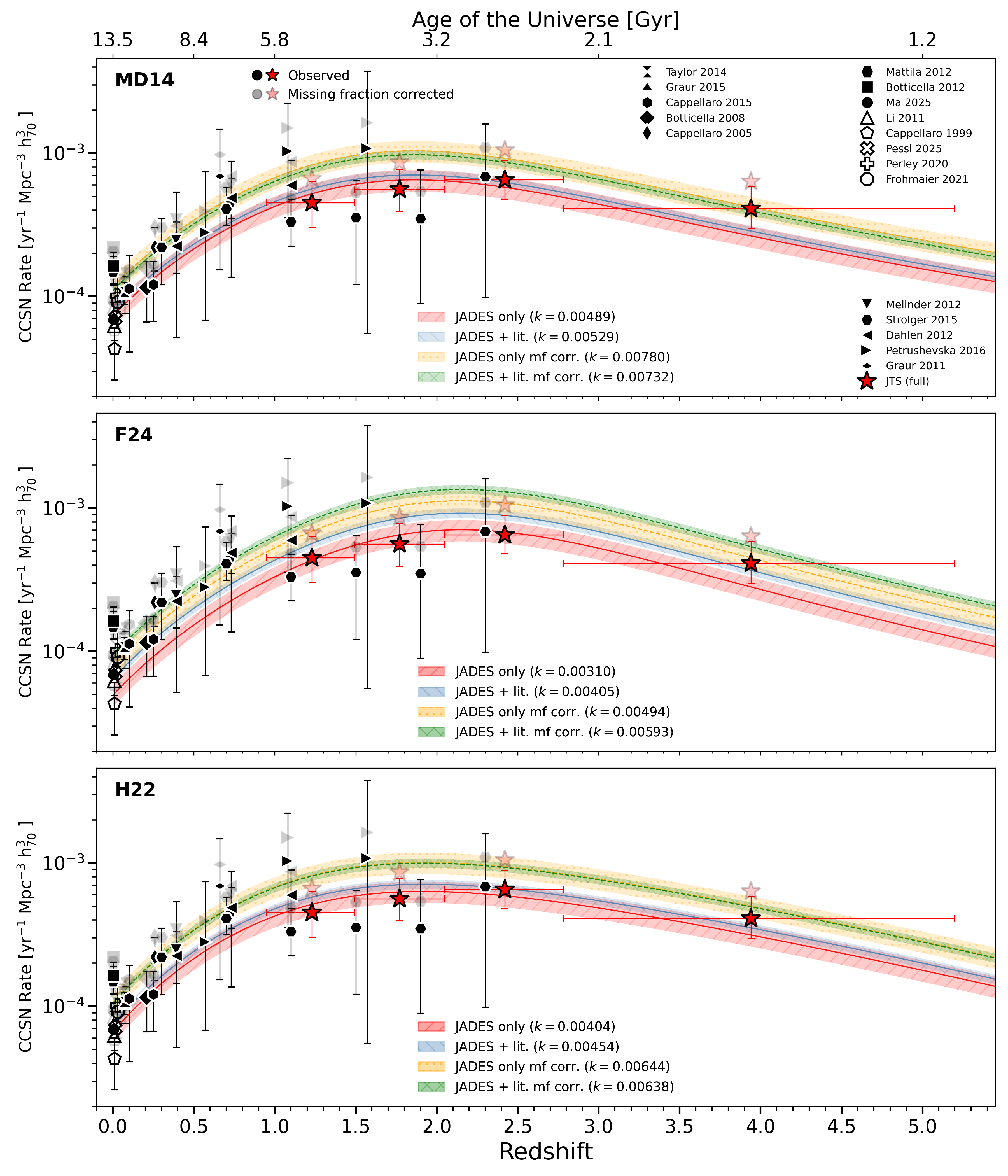}
    \caption{
    Compilation of observed volumetric (comoving volume) CCSN rates from the literature and from the JADES Transient Survey (JTS) (listed in Table~\ref{tab:ccsn_rates}), shown as a function of redshift, the top axis gives the corresponding cosmic age for an $h_{70}$ cosmology. Error bars show the statistical and systematic uncertainties added in quadrature. These measurements are compared to the expected CCSN rate evolution computed from the SFRD of \citet{Madau2014} (top), \citet{Fujimoto24} (middle), and \citet{Harikane22} (bottom), adopting the relation $R_{\rm CCSN}=k_{\rm CCSN} \times \mathrm{SFRD}$. The solid curves show the best-fit $k_{\rm CCSN}$ values obtained by fitting only the JTS CCSN rates (red) and by fitting the full CCSN compilation (blue), with shaded bands indicating the corresponding $1\sigma$ uncertainties. In addition to the observed rates, we show CCSN rates corrected for the nominal missing-supernova fraction (mf) from \citet{Mattila2012a}, plotted as shaded data points. Fits to the corrected JTS-only sample and to the mf corrected full CCSN compilation are shown in yellow and green, respectively. Open symbols indicate CCSN rates that are treated as limits and are not used in any of the fits.
    }
    \label{fig:CCSNe_andk}
\end{figure*}

To quantify the normalization of the CCSN--SFR relation, we determine the
best-fit value of $k_{\mathrm{CCSN}}$ by comparing the expected CCSN rates from
Equation~\ref{eq:ccsn_sfr_relation} with the observed CCSN rate measurements
compiled in Table~\ref{tab:ccsn_rates}. For this analysis, we adopt a Salpeter
IMF and consider parameterized descriptions of the SFRD, the compilation of \citet{Madau2014}, with additional
comparisons using the SFRDs of \citet{Fujimoto24} and \cite{Harikane22}.

The best-fit value of $k_{\mathrm{CCSN}}$ is estimated using a Markov Chain Monte
Carlo (MCMC) approach with flat priors, implemented with the \texttt{emcee}
package \citep{Foreman-Mackey2019}. The likelihood function compares the expected
and observed CCSN rates in each redshift bin,
\begin{equation}
\label{eq:likelihood}
\log \mathcal{L} = -\frac{1}{2}
\sum_i
\left(
\frac{
k_{\mathrm{CCSN}} \times \rho_{\mathrm{SFR}}(z_i)
-
R_{\mathrm{CCSN,obs}}(z_i)
}{
\sigma_{\mathrm{CCSN,obs}}(z_i)
}
\right)^2 ,
\end{equation}
where $R_{\mathrm{CCSN,obs}}$ denotes the observed CCSN rate and
$\sigma_{\mathrm{CCSN,obs}}$ is the total uncertainty, calculated as the average
quadrature sum of the reported statistical and systematic uncertainties listed
in Table~\ref{tab:ccsn_rates}. For the handling of uncertainties in the JADES CCSN rates see C. DeCoursey et al. submitted.

We perform the fits using two CCSN rate samples: (i) the JADES CCSN measurements alone, and (ii) the combined JADES and literature CCSN compilation, excluding measurements that are not corrected for host-galaxy dust extinction (marked with daggers in Table~\ref{tab:ccsn_rates}). Each fit is carried out using both the \citet{Madau2014} and \citet{Fujimoto24} SFRDs, and is repeated after applying a nominal missing-supernova correction to the observed CCSN rates following \citet{Mattila2012a}.

Figure~\ref{fig:CCSNe_andk} shows the observed CCSN rates together with the expected CCSN rate evolution derived from the \citet{Madau2014} (top panel), the \citet{Fujimoto24} (middle panel) and the \citet{Harikane22} (bottom panel) SFRDs. The red and blue curves show the best-fit $k_{\mathrm{CCSN}}$ values obtained using the observed JADES only and combined CCSN samples, respectively, with shaded regions indicating the corresponding $1\sigma$ uncertainties. The nominal missing-supernova correction of \citet{Mattila2012a}, assumed to remain constant beyond $z=2$, is applied to the observed CCSN rates and shown as shaded data points. Fits to the JADES-only sample and to the full CCSN compilation, both corrected for missing supernovae, are shown in yellow and green, respectively.

\begin{figure}[t]
    \centering
    \includegraphics[width=1.05\columnwidth]{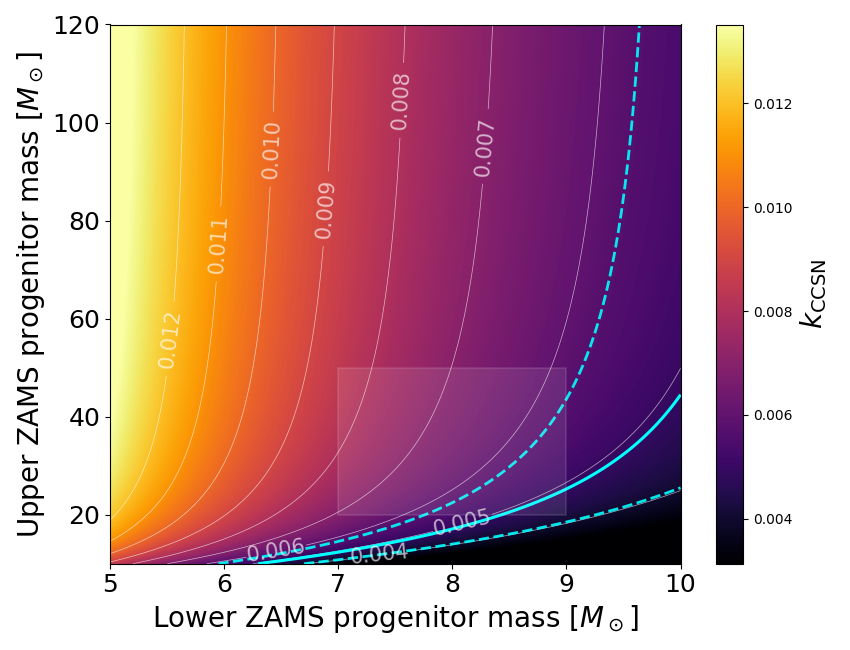}
\caption{
$k_{\mathrm{CCSN}}$ as a function of the lower and upper ZAMS progenitor mass limits, computed using Equation~\ref{eq:ccsn_sfr_relation} and \citet{Salpeter} IMF. Cyan lines mark the best-fit solution (dashed lines indicate the $1\sigma$ uncertainty region) to the JADES-only CCSN rates assuming the SFRD of \citet{Madau2014} and no missing supernova correction. The white square indicates the progenitor mass range broadly compatible with current observational constraints, corresponding to a lower ZAMS mass of $7$--$9\,\mathrm{M_\odot}$ and an upper ZAMS mass of $20$--$50\,\mathrm{M_\odot}$ for CCSN progenitors.}
\label{fig:k_heatmap}
\end{figure}

The resulting best-fit values of $k_{\mathrm{CCSN}}$ are summarized in Table~\ref{tab:k_summary}. Below, we outline the main results and discuss them together with Figure~\ref{fig:k_heatmap}, which shows $k_{\mathrm{CCSN}}$ as a function of the lower and upper ZAMS progenitor-mass limits. All values are computed using Equation~\ref{eq:ccsn_sfr_relation} assuming a Salpeter IMF.

\begin{enumerate}

  \item Madau \& Dickinson (2014, MD14) SFRD

  \begin{enumerate}

  \item \emph{No missing supernova correction.}
  Using the \citet{Madau2014} SFRD without a missing-supernova correction,
  the JADES-only sample yields
  $k_{\mathrm{CCSN}} = 0.0049 \pm 0.0008\,\mathrm{M_\odot^{-1}}$ (cyan lines in Figure~\ref{fig:k_heatmap}),
  while fitting the combined CCSN sample gives a consistent value of
  $k_{\mathrm{CCSN}} = 0.0053 \pm 0.0004\,\mathrm{M_\odot^{-1}}$.

  \item \emph{With missing supernova correction.}
  Applying the missing-supernova correction to the JADES-only sample yields
  a higher efficiency of $k_{\mathrm{CCSN}} = 0.0078 \pm 0.0014\,\mathrm{M_\odot^{-1}}$. Applying the same correction to the combined CCSN sample gives a slightly lower efficiency of $k_{\mathrm{CCSN}} = 0.0073 \pm 0.0005\,\mathrm{M_\odot^{-1}}$, which for a lower ZAMS mass limit of $8\,\mathrm{M_\odot}$ corresponds to an upper limit of $\sim$$86\,\mathrm{M_\odot}$ ($\gtrsim 40\,\mathrm{M_\odot}$ within $1\sigma$), i.e. a wide progenitor mass range.

  \end{enumerate}

  \item Fujimoto et al. (2024, F24) SFRD

  \begin{enumerate}

  \item \emph{No missing supernova correction.}
  Using the \citet{Fujimoto24} SFRD without a missing-supernova correction yields systematically lower efficiencies. For the JADES-only sample, we obtain $k_{\mathrm{CCSN}} = 0.0031 \pm 0.0005\,\mathrm{M_\odot^{-1}}$. For the combined CCSN sample, $k_{\mathrm{CCSN}} = 0.0041 \pm 0.0003\,\mathrm{M_\odot^{-1}}$.

  \item \emph{With missing supernova correction.}
  Applying the missing-supernova correction increases the inferred efficiency to
  $k_{\mathrm{CCSN}} = 0.0049 \pm 0.0008\,\mathrm{M_\odot^{-1}}$ for the JADES-only sample and $k_{\mathrm{CCSN}} = 0.0059 \pm 0.0004\,\mathrm{M_\odot^{-1}}$ for the combined CCSN sample.

  \end{enumerate}

  \item Harikane et al. (2022, H22) SFRD

  \begin{enumerate}

  \item \emph{No missing supernova correction.}
  Using the \citet{Harikane22} SFRD without a missing-supernova correction
  yields
  $k_{\mathrm{CCSN}} = 0.0040 \pm 0.0007\,\mathrm{M_\odot^{-1}}$
  for the JADES-only sample and
  $k_{\mathrm{CCSN}} = 0.0045 \pm 0.0003\,\mathrm{M_\odot^{-1}}$
  for the combined CCSN sample.

  \item \emph{With missing supernova correction.}
  Applying the missing-supernova correction increases the efficiency to
  $k_{\mathrm{CCSN}} = 0.0064 \pm 0.0011\,\mathrm{M_\odot^{-1}}$
  for the JADES-only sample and
  $k_{\mathrm{CCSN}} = 0.0064 \pm 0.0005\,\mathrm{M_\odot^{-1}}$
  for the combined CCSN sample.

  \end{enumerate}

\end{enumerate}

\begin{table}[ht]
\centering
\caption{Summary of $k_{\mathrm{CCSN}}$ fits used in Section \ref{sec:LinkingSFR}}
\label{tab:k_summary}
\begin{tabular}{lll}
\hline
\hline
$k_{\mathrm{CCSN}}~[\mathrm{M_\odot^{-1}}]$ & Notes \\
\hline
 $0.0053 \pm 0.0004$ & MD14 SFRD, all data \\
 $0.0049 \pm 0.0008$ & MD14 SFRD, JADES data \\
 $0.0073 \pm 0.0005$ & MD14 SFRD, all data, mf \\
  $0.0078 \pm 0.0014$ & MD14 SFRD, JADES data, mf \\
  \hline
 $0.0041 \pm 0.0003$ & F24 SFRD, all data \\
 $0.0031 \pm 0.0005$ & F24 SFRD, JADES data \\
 $0.0059 \pm 0.0004$ & F24 SFRD, all data, mf \\
 $0.0049 \pm 0.0008$ & F24 SFRD, JADES data, mf \\
  \hline
 $0.0045 \pm 0.0003$ & H22 SFRD, all data \\
 $0.0040 \pm 0.0007$ & H22 SFRD, JADES data \\
 $0.0064 \pm 0.0005$ & H22 SFRD, all data, mf \\
 $0.0064 \pm 0.0011$ & H22 SFRD, JADES data, mf \\
\hline
\end{tabular}
{\footnotesize
$k_{\mathrm{CCSN}}$ is the number of CCSNe per unit stellar mass formed, expressed in $\mathrm{M_\odot^{-1}}$. Fixed values are computed assuming a Salpeter IMF. MD14, F24, and H22 denote the SFRDs of \citet{Madau2014}, \citet{Fujimoto24}, and \citet{Harikane22}, respectively. mf indicates that the observed CCSN rates have been corrected for the nominal missing supernova fraction following \citet{Mattila2012a}.
}
\end{table}

In summary, the combination most difficult to reconcile with plausible CCSN progenitor-mass limits within the simplified framework of Equation~\ref{eq:ccsn_sfr_relation} is the uncorrected CCSN rates together with the higher \citet{Fujimoto24} (F24) SFRD, which yields systematically low efficiencies ($k_{\mathrm{CCSN}} \approx 0.0031$--$0.0040\,\mathrm{M_\odot^{-1}}$) and would require implausibly high lower ZAMS mass limits ($\gtrsim 10\,\mathrm{M_\odot}$, Figure~\ref{fig:k_heatmap}). However, applying the missing-SN correction yields $k_{\mathrm{CCSN}} \approx 0.0049$--$0.0059\,\mathrm{M_\odot^{-1}}$ corresponding to progenitor ZAMS mass ranges in agreement with expectations.
Fits to the \citet{Madau2014} (MD14) and \citet{Harikane22} (H22) SFRDs yield efficiencies ranging from $k_{\mathrm{CCSN}} \approx 0.0040$--$0.0053\,\mathrm{M_\odot^{-1}}$ when not correcting for missing SNe to $k_{\mathrm{CCSN}} \approx 0.0064$--$0.0078\,\mathrm{M_\odot^{-1}}$ when correction has been applied. If assuming a lower ZAMS progenitor mass of $8\,\mathrm{M_\odot}$ these efficiencies correspond to upper ZAMS progenitor masses in the $14$--$19\,\mathrm{M_\odot}$ and $30$--$120\,\mathrm{M_\odot}$ ranges, respectively. This highlights that a consistent correction for the effects of dust obscuration for both the SFRDs and the CCSN rates is crucial for the progenitor mass ranges to be more reliably restricted.

\subsection{Effects of the IMF on Expected CCSN Rates}
\label{sec:systematics2}

\begin{figure}
    \centering
    \includegraphics[width=\linewidth]{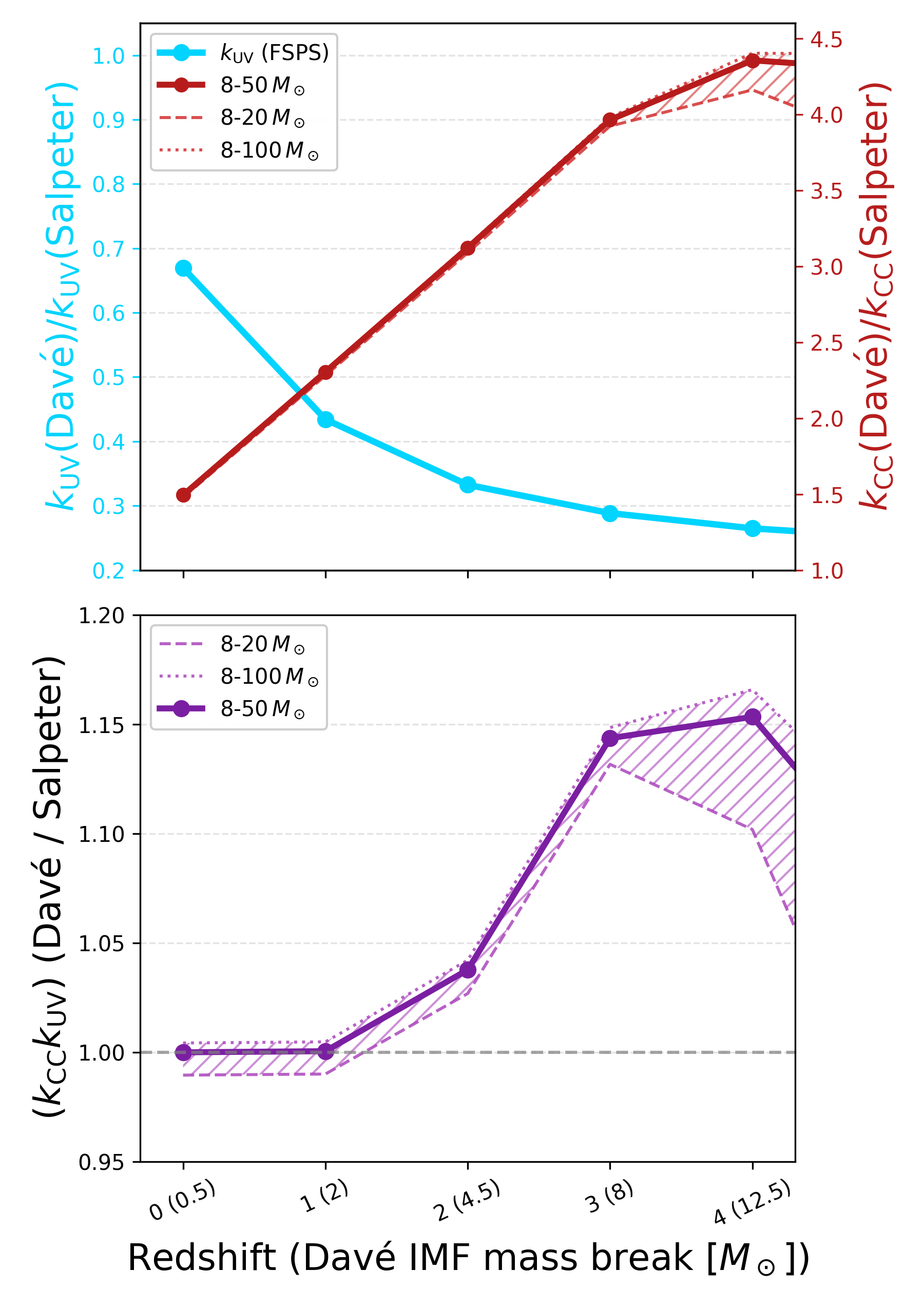}
    \caption{Impact of the Davé IMF on $k_{\rm UV}$ and $k_{\rm CCSN}$ relative to a Salpeter IMF. Top: combined $k_{\rm UV}$ (left axis) and $k_{\rm CC}$ (right axis) ratios relative to Salpeter as a function of redshift. Bottom: the product $(k_{\rm CCSN} \times k_{\rm UV})$ relative to Salpeter. In both panels the hatched region spans the $8$--$20$ to $8$--$100\,\mathrm{M_\odot}$ progenitor range, with the $8$--$50\,\mathrm{M_\odot}$ as solid line.}
    \label{fig:persal_panels}
\end{figure}

The value of $k_{\mathrm{CCSN}}$ is also affected by the adopted IMF, however, a change in the IMF also affects the SFRDs estimated from the galaxy luminosities because massive stars dominate the SFRDs due to their high luminosities, and therefore the effects of changing the IMF mostly cancel out \citep[e.g.][]{DahlenFransson, Dahlen2012a, Melinder2012}. For standard IMFs (\citep[e.g.][]{Salpeter, Kroupa, Chabrier2003}) the effect on the CCSN - SFRD relation is only a few percent.

We are also interested in testing the effects of a more top-heavy IMF and a redshift evolving IMF. For this analysis, we adopt the redshift-dependent IMF model of \citet{Davee2008}, which describes a set of possible IMF shapes using a characteristic mass (mass at which the IMF turns over) that evolves with redshift. Within this framework, the relation between the CCSN rate and the SFRD is written in Equation \ref{eq:ccsn_sfr_relation}, where $k_{\rm CCSN}$ is the CCSN production efficiency per unit stellar mass formed. The SFR is related to the rest-frame UV luminosity through
\begin{equation}
\label{eq:kuv}
\rho_{\mathrm{SFR}} = k_{\rm UV}(\mathrm{IMF}) \times L_{\rm UV},
\end{equation}

where the UV conversion factor $k_{\rm UV}$ depends on the assumed IMF (\citealt{kennicutt1998, Madau2014}). Combining Equations \ref{eq:ccsn_sfr_relation} and \ref{eq:kuv}, the expected CCSN rate can be written as
\begin{equation}
R_{\rm CCSN}(\mathrm{IMF}) = k_{\rm CCSN}(\mathrm{IMF}) \times k_{\rm UV}(\mathrm{IMF}) \times L_{\rm UV}.
\end{equation}

The \cite{Davee2008} IMF is parameterized by a characteristic mass that is dependent on redshift, which leads to a progressively more top-heavy IMF with increasing redshift.
Although this IMF prescription is calibrated using observations only up to $z \sim 2$, we apply it to higher redshifts to explore the potential impact of IMF evolution on the CCSN rate–SFRD relation in the absence of alternative observationally constrained models.

To calculate $k_{\mathrm{CCSN}}$, we insert the chosen IMF into Equation \ref{eq:ccsn_sfr_relation}, assuming the CCSN progenitor mass limits of 8–20~$\mathrm{M_\odot}$, 8–50~$\mathrm{M_\odot}$ and 8–100~$\mathrm{M_\odot}$. The IMF dependence of the SFRD enters through the conversion factor $k_{\rm UV}$, which relates the measured UV luminosities to the SFRs.
\citet{Madau2014} report conversion factors for SFRD assuming the Kroupa and Chabrier IMFs of 0.67 and 0.63, respectively, compared to SFRD assuming the Salpeter IMF.

Here, we have used the same stellar population synthesis model as in \citet{Madau2014}, the \texttt{FSPS} code \citep{FSPS}, to determine $k_{\rm UV}$ for the \cite{Davee2008} IMFs. As a consistency check, we verified that our \texttt{FSPS} simulations reproduce the \citet{Madau2014} result for $k_{UV}$ with a Salpeter IMF. Our results agree within 3\% with those given in \citet{Madau2014}.

The \texttt{FSPS} setup follows \citet{Madau2014}, assuming a constant SFR for calculating $k_{\rm UV}$ for several metallicities
$\left(\log \frac{Z_*}{Z_\odot} = +0.2,\, 0,\, -0.5,\, -1.0 \right)$
and a stellar population age of $\sim 300$~Myr.
We used the 1500\,\AA\ luminosity to determine the $k_{\rm UV}$ values and averaged them over metallicity.
This procedure is performed with the Salpeter IMF and is repeated for the Davé IMFs, with stellar mass cutoffs of 0.1–100~$\mathrm{M_\odot}$. The simulation was performed using identical setups for the Salpeter IMF and the \cite{Davee2008} IMFs.
The results are shown in Figure~\ref{fig:persal_panels}: the top panel (cyan line) presents $k_{\rm UV}$ relative to its value assuming the Salpeter IMF, while $k_{\mathrm{CCSN}}$ (Equation \ref{eq:ccsn_sfr_relation}) is calculated relative to Salpeter for three different maximum progenitor masses, indicated by the uncertainty band.
The bottom panel shows the combined impact of the IMF variation in $k_{\mathrm{CCSN}}$ and $k_{\rm UV}$ on the expected CCSN rate. The results suggest that the effect on the CCSN - SFRD relation is negligible given the current uncertainties in both CCSN rates and SFRDs. In particular, within $z<2$, the redshift range over which the \cite{Davee2008} IMF is calibrated against observations, the combined effect on the CCSN - SFRD relation remains below 5\%.

This behavior can be understood as a consequence of the coupled role of the IMF both in setting the CCSN production efficiency ($k_{\mathrm{CCSN}}$) and in the inferred SFRD. While a more top-heavy IMF increases the value of $k_{\mathrm{CCSN}}$ by enhancing the fraction of massive, CCSN producing stars per unit stellar mass, it simultaneously affects the luminosity based SFRD estimates. More massive stars are substantially more luminous, with stellar luminosity scaling approximately as $L \propto M^{2\text{--}3}$. For a fixed observed luminosity density, an IMF weighted toward higher stellar masses therefore requires a smaller total stellar mass to reproduce the same luminosity, leading to a reduced inferred SFRD. These two effects largely cancel out in the CCSN rate-SFRD relation.

At early cosmic times (z $>$ 6), it has been suggested that stellar populations may have followed an extremely top-heavy IMF, potentially leading to the production of pair-instability supernovae (PISNe). Several studies propose physically motivated IMF forms appropriate for such environments (e.g., \citealt{venditti}). An early Universe IMF introduced by \cite{venditti} is parameterized as
\begin{equation}
\phi(m) \propto m^{-a}\,\exp\left[-\left(\frac{m_c}{m}\right)^b\right],
\end{equation}
with $a = 2.35$, $b = 1$, and $m_c = 10\,M_\odot$.

Using this IMF over a stellar mass range of $1$--$500\,M_\odot$, and assuming SN progenitors between $8$ and $260\,M_\odot$ (thus including the PISN mass range), we obtain according to equation \ref{eq:ccsn_sfr_relation}:
\begin{equation}
k_{\mathrm{CCSN}} = 0.028\,M_\odot^{-1}.
\end{equation}

The \texttt{FSPS} framework does not include this specific IMF form, preventing a direct calculation of the corresponding $k_{\mathrm{UV}}$. However, another top-heavy IMF suggested in \cite{harikane23} (for redshifts 9 $<$ z $<$ 16), namely their PopIII.2 log-normal IMF model, is used here. In \cite{harikane23} Figure 20, they report
\begin{equation}
k_{\mathrm{UV}} = 0.40 \times 10^{-28}\, 
M_\odot\,\mathrm{yr}^{-1}\,\mathrm{erg}^{-1}\,\mathrm{s}\,\mathrm{Hz},
\end{equation}

and we calculate according to equation \ref{eq:ccsn_sfr_relation}:
\begin{equation}
k_{\mathrm{CCSN}} = 0.013\,M_\odot^{-1}.
\end{equation}

These values imply that switching from a Salpeter IMF even to such an extreme top-heavy IMF would only change the expected CCSN rate by approximately $\sim 35\%$. Although this effect is larger than those found for the IMF variations that we tested at lower redshifts, it would still remain undetected given the large uncertainties in the current CCSN rates estimated at the highest redshifts and other uncertainties including progenitor mass ranges and dust obscuration. Also, for the redshift range considered in this work ($z = 0$--5), such extreme IMFs are unlikely to be relevant and, therefore, we do not consider them in our primary analysis. According to this analysis, constraining the IMF from CCSN rates alone seems challenging, and other information such as relative rates between different CCSN subtypes (e.g., \citealt{Graur2017_2}) or alternative methods (e.g., \citealt{fraser2017}) would be required.

\section{Constraints on the Cosmic Star-Formation History from CCSNe}
\label{sec:constrains}

For the analyzes in this Section and in Section \ref{sec:MF} we use the JADES CCSN rates together with a compilation of literature measurements at $z<1$ as the observed CCSN rates, to anchor the low-redshift range. In this way, the primary focus is on the JADES sample, and the low-redshift data provide essential constraints. In this Section we apply no missing supernova fraction on the observed CCSN rates.

\subsection{CCSN-derived Cosmic SFRD}
\label{sec:constrains2}

In addition to predicting CCSN rates from an assumed SFRD, the observed CCSN rates can be used to
constrain the cosmic SFRDs directly. This
approach can provide an independent probe of the CSFH that is complementary to the galaxy luminosity-based
methods.

We model the SFRD using the functional form introduced by
\citet{Madau2014},
\begin{equation}
\label{eq:sfr_params}
\rho_{\mathrm{SFR}}(z)=
\frac{A(1+z)^B}{1+\left[(1+z)/C\right]^D},
\end{equation}

where $A$, $B$, $C$, and $D$ are free parameters. In this analysis, the CCSN--SFRD conversion factor $k_{\mathrm{CCSN}}$ is fixed. For this we adopt three representative values, $k_{\mathrm{CCSN}} = 0.0054$, $0.0070$, and $0.0089\,\mathrm{M_\odot^{-1}}$, corresponding to CCSN progenitor mass ranges of $8$--$20\,\mathrm{M_\odot}$, $8$--$50\,\mathrm{M_\odot}$, and $7$--$120\,\mathrm{M_\odot}$, respectively. We use these as an example, and other progenitor ZAMS mass ranges are possible. The expected CCSN rates are computed as in Equation \ref{eq:ccsn_sfr_relation} and compared with the observed CCSN rates listed in Table~\ref{tab:ccsn_rates}. The likelihood function is defined as in Equation \ref{eq:likelihood}.

Flat priors are adopted for all four SFRD parameters,
with ranges
$A \in [0,0.4]$,
$B \in [0,7]$,
$C \in [0,7]$, and
$D \in [0,10]$.

\begin{figure}[htb]
\centering
\includegraphics[width=0.49\textwidth]{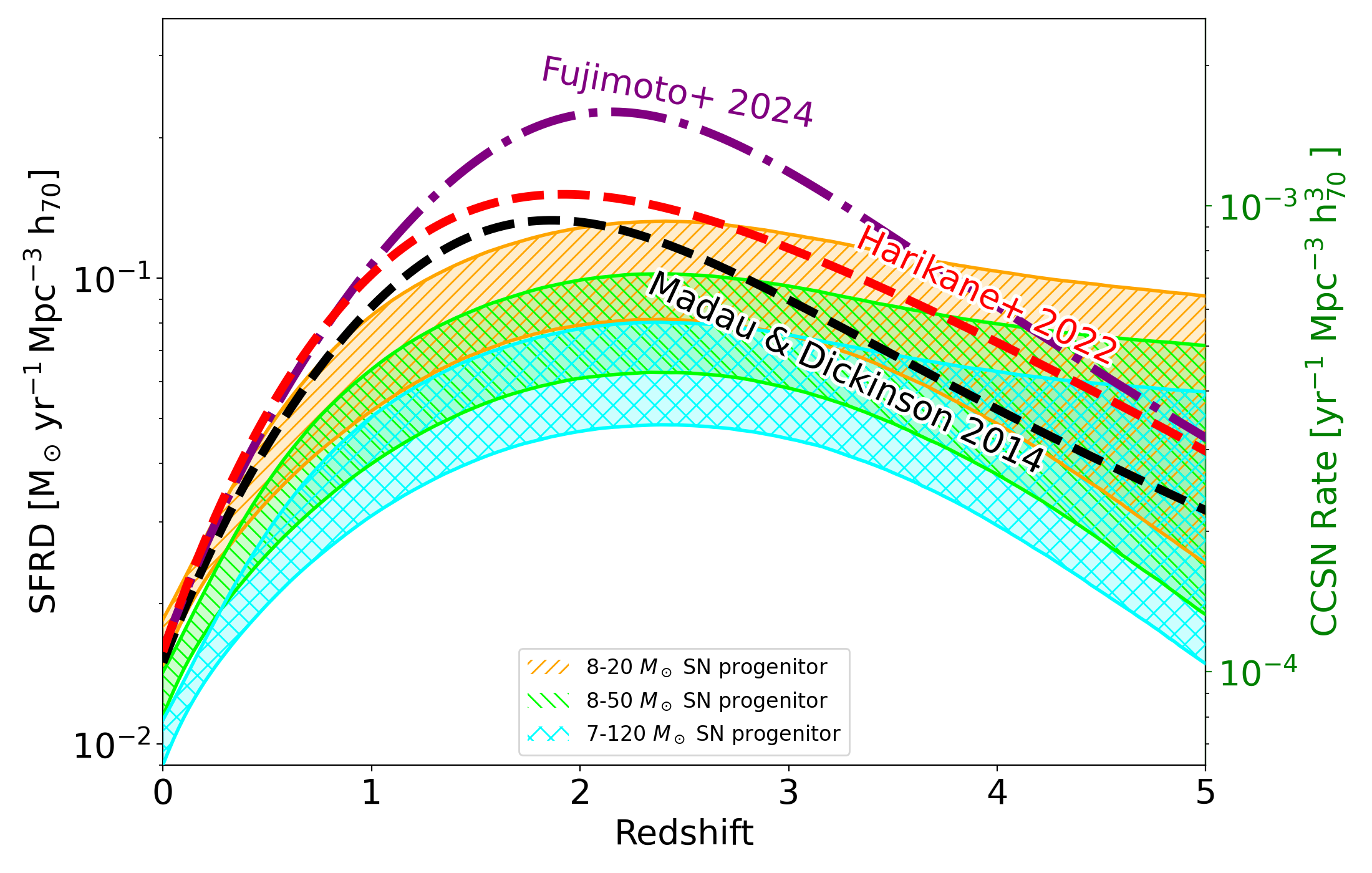}
\caption{
SFRDs inferred from the CCSN rate measurements for three fixed values of the CCSN efficiency factor $k_{\mathrm{CCSN}}$, corresponding to progenitor mass ranges of $8$–$20\,\mathrm{M_\odot}$ (orange), $8$–$50\,\mathrm{M_\odot}$ (green), and $7$–$120\,\mathrm{M_\odot}$ (cyan). Shaded regions indicate the $1\sigma$ credible intervals. For reference, SFRDs of \citet{Madau2014}, \citet{Harikane22} and \citet{Fujimoto24} are shown as dashed curves. The latter is converted to a Salpeter IMF following the conversion guidelines of \citet{Madau2014}. The right-hand axis shows the CCSN rate corresponding to $\mathrm{SFRD}\times0.0070\mathrm{M_\odot^{-1}}$ ($8$–$50\,\mathrm{M_\odot}$ progenitors).}
\label{fig:SFRDs}
\end{figure}

The resulting CCSN-derived SFRDs are shown in Fig.~\ref{fig:SFRDs} for the three adopted values of $k_{\mathrm{CCSN}}$. The corresponding corner plot and the resulting fitted SFRD parameters are presented in
Appendix Figure \ref{fig:bestsfrs} and Appendix Table \ref{tab:csfh_zams_summary} respectively. For the case $k_{\mathrm{CCSN}} = 0.0070\,M_{\odot}^{-1}$ the fitted SFRD parameters are:
\begin{align*}
A &= 0.013^{+0.002}_{-0.002}\,[\,\mathrm{M_\odot}\,\mathrm{yr}^{-1}\,\mathrm{Mpc}^{-3}\,], \nonumber\\
B &= 2.2^{+1.0}_{-0.5}, \nonumber\\
C &= 3.3^{+1.1}_{-1.3}, \nonumber\\
D &= 4.9^{+2.3}_{-1.4}. \label{eq:sfrd_bestfit_params}
\end{align*}

For $k_{\mathrm{CCSN}} = 0.0054\,M_{\odot}^{-1}$ only the normalization changes, with $A = 0.017^{+0.002}_{-0.002}$ and $B$, $C$, $D$ unchanged, since $k{_\mathrm{CCSN}}$ rescales only the overall amplitude.
 
The redshifts at which the SFRDs peak are determined for each posterior sample individually, constructing the corresponding distribution of peak redshifts and determining the median and central 68\% interval. Using all posterior samples, we obtain $z_{\rm peak}=2.4^{+1.0}_{-0.6}$, independent of the adopted $k_{\mathrm{CCSN}}$. The large upper uncertainty arises from a tail of $\sim$10\% of the samples that peak beyond $z=5$. Since our CCSN rate measurements extend only to $z\approx5$, the peak positions of these samples are unconstrained by the data. Excluding them, we obtain $z_{\rm peak}=2.3^{+0.6}_{-0.5}$. both determinations are consistent within $1\sigma$ with galaxy luminosity-based determinations of the SFRD, such as \citet{Madau2014}, who report a peak at $z \sim 1.9$.

\subsection{Posterior Predictive Statistical Tests}
\label{sec:constrains3}

\begin{table}
  \centering
  \caption{
  Posterior predictive test statistics and hypothesis test results for the null
  hypothesis that the selected SFRD curves are consistent with the
  posterior predictive distribution of our model, for the SFRDs of
  \citet{Madau2014} (MD14), \citet{Harikane22} (H22), and \citet{Fujimoto24}
  (F24). For each statistic we list its value, the posterior predictive
  $p$-value, and whether the null hypothesis is rejected at the 68\% / 95\% / 99\% levels
  (i.e. $p < 0.32$, $0.05$, $0.01$). In the final column, $0$ denotes rejection that the SFRD is consistent with our model and $1$ failure to reject, at the 68\% / 95\% / 99\% levels respectively. Results are listed for two efficiencies, $k_{\mathrm{CCSN}} = 0.0070$ and $0.0054\,\mathrm{M_\odot^{-1}}$.
  \label{tab:posterior_tests}
  }
  \begin{tabular}{llcccc}
  \hline
  \hline
  Stat. & SFRD & $k_{\mathrm{CCSN}}$ & Value & $p$ & 68/95/99\% \\
  \hline
  \multirow{6}{*}{RMS}
    & \multirow{2}{*}{MD14} & 0.0070 & 0.16 & 0.25 & 0 / 1 / 1 \\
    &                       & 0.0054 & 0.12 & 0.44 & 1 / 1 / 1 \\
    \cline{2-6}
    & \multirow{2}{*}{H22} & 0.0070 & 0.21 & 0.09 & 0 / 1 / 1 \\
    &                       & 0.0054 & 0.12 & 0.46 & 1 / 1 / 1 \\
    \cline{2-6}
    & \multirow{2}{*}{F24} & 0.0070 & 0.31 & 0.03 & 0 / 0 / 1 \\
    &                       & 0.0054 & 0.21 & 0.10 & 0 / 1 / 1 \\
  \hline
  \multirow{6}{*}{KS}
    & \multirow{2}{*}{MD14} & 0.0070 & 0.12 & 0.23 & 0 / 1 / 1 \\
    &                       & 0.0054 & 0.12 & 0.24 & 0 / 1 / 1 \\
    \cline{2-6}
    & \multirow{2}{*}{H22} & 0.0070 & 0.084 & 0.41 & 1 / 1 / 1 \\
    &                       & 0.0054 & 0.083 & 0.43 & 1 / 1 / 1 \\
    \cline{2-6}
    & \multirow{2}{*}{F24} & 0.0070 & 0.097 & 0.33 & 1 / 1 / 1 \\
    &                       & 0.0054 & 0.095 & 0.35 & 1 / 1 / 1 \\
  \hline
  \multirow{6}{*}{$\chi^2$}
    & \multirow{2}{*}{MD14} & 0.0070 & 510 & 0.08 & 0 / 1 / 1 \\
    &                       & 0.0054 & 160 & 0.48 & 1 / 1 / 1 \\
    \cline{2-6}
    & \multirow{2}{*}{H22} & 0.0070 & 980 & 0.02 & 0 / 0 / 1 \\
    &                       & 0.0054 & 280 & 0.24 & 0 / 1 / 1 \\
    \cline{2-6}
    & \multirow{2}{*}{F24} & 0.0070 & 1800 & $<0.01$ & 0 / 0 / 0 \\
    &                       & 0.0054 & 790 & 0.03 & 0 / 0 / 1 \\
  \hline
  \end{tabular}
\end{table}

To quantitatively assess the consistency between the CCSN-derived and galaxy luminosity based SFRDs, we apply posterior predictive statistical tests following the MCMC analysis. This analysis is performed using the $k_{\mathrm{CCSN}} = 0.0070\,\mathrm{M_\odot^{-1}}$ and $0.0054\,\mathrm{M_\odot^{-1}}$ cases (green and orange areas in Figure \ref{fig:SFRDs}, respectively).
 
For each posterior draw of the SFRD parameters, we compute three statistics relative to the median posterior model: the root-mean-square (RMS) offset, the Kolmogorov--Smirnov (KS) statistic, and the $\chi^2$ statistic. These statistics characterize differences in overall normalization (RMS), shape (KS), and pointwise deviations weighted by posterior variance ($\chi^2$).

The RMS statistic is computed as the quadratic mean of the difference between each posterior draw and the median posterior curve. The KS statistic is defined as the maximum absolute difference between the cumulative distribution functions of the posterior draw and the median posterior created by dividing each curve by its integral and cumulatively summing the result over redshift. The $\chi^2$ statistic is calculated using the posterior variance at each redshift as the uncertainty.

We then compute the same statistics for the galaxy luminosity based SFRDs by comparing them to the median posterior curve. The resulting values are placed within the posterior-derived distributions to determine one sided $p$-values, representing how likely our posterior model is to produce a result as extreme as the selected SFRD, where a low $p$-value indicates poor agreement.

Figure~\ref{fig:stat_tests} shows the posterior predictive distributions for all three statistics, with the literature SFRD curves indicated for the two conversion factors $k_{\mathrm{CCSN}} = 0.0070$ and $0.0054,\mathrm{M_\odot^{-1}}$ ($8$--$50$ and $8$--$20,\mathrm{M_\odot}$ progenitor mass ranges, respectively). The numerical results and hypothesis-test outcomes for both values are summarized in Table~\ref{tab:posterior_tests}.

The \citet{Madau2014} SFRD is broadly consistent with the CCSN-derived posterior at both efficiencies. For $k_{\mathrm{CCSN}} = 0.0070\,\mathrm{M_\odot^{-1}}$ the RMS ($p=0.25$), KS ($p=0.23$), and $\chi^2$ ($p=0.078$) statistics each indicate only marginal 68\% tension, while for $0.0054\,\mathrm{M_\odot^{-1}}$ only the KS test retains 68\% tension ($p=0.24$), with the RMS ($p=0.44$) and $\chi^2$ ($p=0.48$) tests consistent.

The \citet{Harikane22} SFRD shows a slightly worse agreement. For $k_{\mathrm{CCSN}} = 0.0070\,\mathrm{M_\odot^{-1}}$ the KS test does not reject the null hypothesis ($p=0.41$), while the RMS ($p=0.095$) and $\chi^2$ ($p=0.022$) statistics reject it at the 68\% level, with the $\chi^2$ test also rejecting at 95\%, for $0.0054\,\mathrm{M_\odot^{-1}}$ the tension reduces to a marginal 68\% rejection by the $\chi^2$ test ($p=0.24$).

The \citet{Fujimoto24} SFRD shows the strongest tension at both efficiencies. For $k_{\mathrm{CCSN}} = 0.0070\,\mathrm{M_\odot^{-1}}$ the KS test does not reject the null hypothesis ($p=0.33$), but the RMS statistic rejects it at the 68\% and 95\% levels ($p=0.033$) and the $\chi^2$ test rejects it at all three levels ($p=0.007$), for $0.0054\,\mathrm{M_\odot^{-1}}$ this tension softens, with the RMS statistic rejecting only at the 68\% level ($p=0.10$) and the $\chi^2$ test rejecting at the 68\% and 95\% levels ($p=0.034$) but no longer at 99\%.

In summary, the agreement worsens from \citet{Madau2014} through \citet{Harikane22} to \citet{Fujimoto24}, tracking their increasing SFRDs: RMS and $\chi^2$ tests progressively reject the higher SFRDs, whereas the shape sensitive KS test shows at most marginal 68\% tension for any model. The tension therefore reflects the overall SFRD normalization rather than its redshift shape, and it softens for the lower efficiency $k_{\mathrm{CCSN}} = 0.0054,\mathrm{M_\odot^{-1}}$, where no model is rejected at the 99\% level. In general, our CCSN-derived SFRDs favor the lower \citet{Madau2014} SFRD. This could be seen as an indication that CCSN surveys trace the same star formation as the SFRD studies relying on galaxy UV luminosities at high redshift and that the dust extinction corrections used are comparable.

\section{Supernovae missed due to dust obscuration}
\label{sec:MF}

We implemented another MCMC analysis to constrain the redshift dependent missing supernova fraction that affects the observed CCSN rates. The expected CCSN rates are computed using Equation \ref{eq:ccsn_sfr_relation}, adopting the SFRD of \citet{Fujimoto24}.

For this analysis, we adopt a wide range of different progenitor mass limits and their resulting $k_{\mathrm{CCSN}}$ values, which are presented in Table \ref{tab:mf_zams_summary} first and second columns. The SFRD is from \citet{Fujimoto24} and converted to a Salpeter IMF normalization by dividing by 0.63, following \citet{Madau2014}.

\begin{figure}[hbt]
\centering
    \centering
    \includegraphics[width=0.49\textwidth]{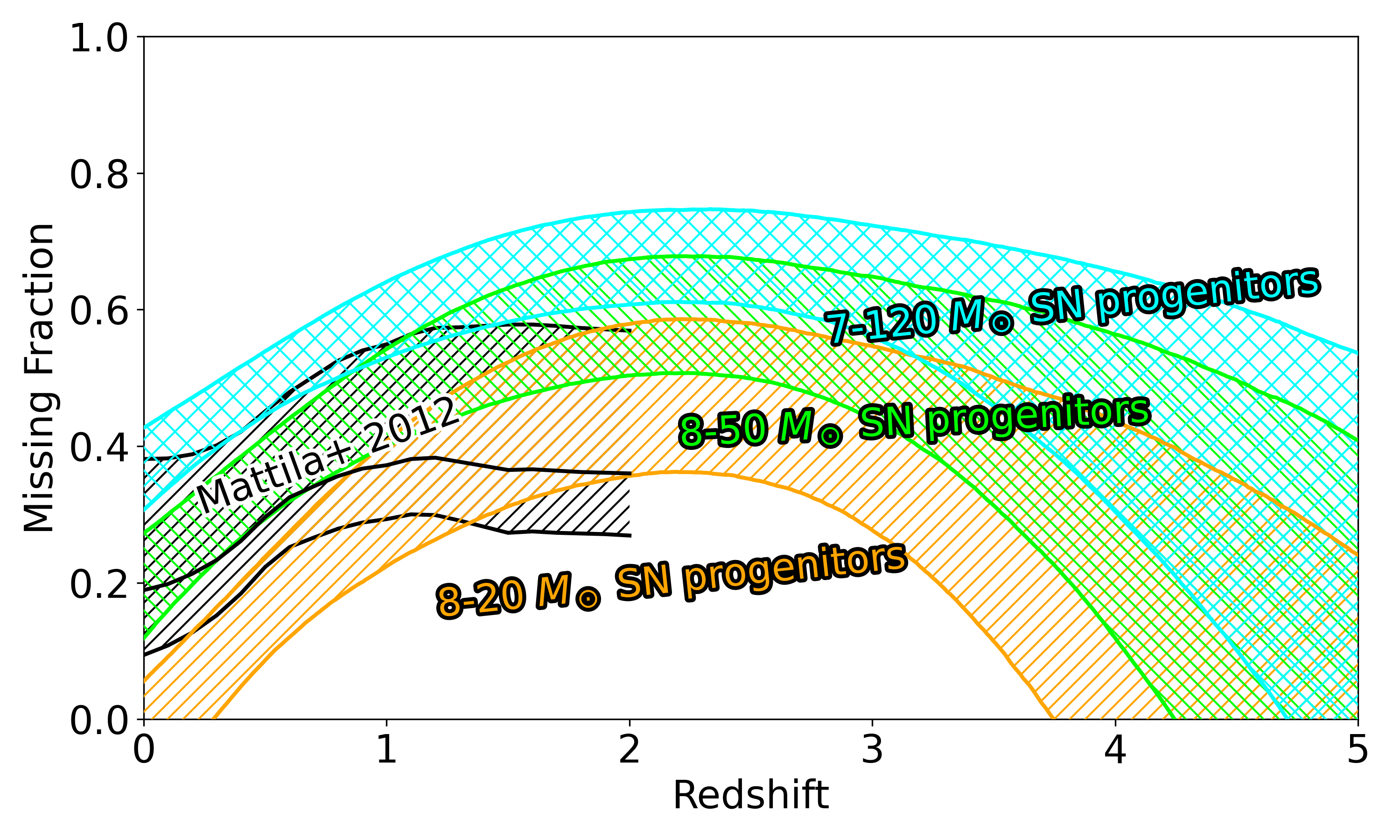}
    \caption{Comparison of the missing supernova fractions inferred using second degree polynomial models. The model CCSN rate is computed using the \citet{Fujimoto24} SFRD, converted from a Chabrier to a Salpeter IMF by dividing by 0.63. Three progenitor mass ranges are adopted to define the CCSN–SFRD conversion factor $k_{\mathrm{CCSN}}$: $8$--$20\,\mathrm{M_\odot}$, $8$--$50\,\mathrm{M_\odot}$, and $7$--$120\,\mathrm{M_\odot}$, corresponding to $k_{\mathrm{CCSN}} = 0.0054$, $0.0070$, and $0.0089\,\mathrm{M_\odot^{-1}}$, respectively, as given by Equation~\ref{eq:ccsn_sfr_relation}. The missing supernova fraction derived by \citet{Mattila2012a} up to $z=2$ is shown in black.}
\label{fig:mf_poly_compare}
\end{figure}

The adopted CCSN rate sample for this analysis is the same sample used in Section \ref{sec:constrains}. To account for CCSNe missed due to dust obscuration, a redshift-dependent missing fraction correction is applied to the observed rates and their uncertainties. The correction factor is defined as \begin{equation} C(z) = \frac{1}{1 - f_{\mathrm{miss}}(z)} , \end{equation} such that the corrected CCSN rates and uncertainties are given by $R_{\mathrm{CCSN,corrected}} = C \times R_{\mathrm{CCSN,observed}}$ and $\sigma_{\mathrm{CCSN,corrected}} = C \times \sigma_{\mathrm{CCSN,observed}}$. The log-likelihood function minimizes the chi-square difference between the expected and observed CCSN rates and is written as
\begin{equation}
\label{eq:loglike}
\begin{aligned}
\log \mathcal{L}
&= -\frac{1}{2}\sum_i
\left[
\frac{
k_{\mathrm{CCSN}}\,\rho_{\mathrm{SFR}}(z_i)
- C(z_i)\,R_{\mathrm{CCSN,obs}}(z_i)
}{
C(z_i)\,\sigma_{\mathrm{CCSN,obs}}(z_i)
}
\right]^2 .
\end{aligned}
\end{equation} The uncertainty $\sigma_{\mathrm{CCSN,observed}}$ is the total uncertainty of each CCSN rate measurement, as defined in Section~\ref{sec:LinkingSFR3}. The redshift dependent missing SN fraction is modeled as a continuous second order polynomial, \begin{equation}\label{eq:MF_params} f_{\mathrm{miss}}(z) = f_1 + f_2\,z + f_3\,z^2 , \end{equation} where the parameters $(f_1, f_2, f_3)$ are inferred through MCMC sampling. Flat priors are adopted, with all parameters constrained to within $-1 \leq f_1,\, f_2,\, f_3 \leq 1$. The MCMC analysis uses 32 walkers to explore the three-dimensional parameter space over 50\,000 steps per walker. The initial walker positions are drawn randomly from within the prior range. The resulting missing SN fractions are shown in Figure \ref{fig:mf_poly_compare}. For simplicity we only present three progenitor mass range choices. The three progenitor mass ranges are $8$–$20~\mathrm{M_\odot}$, $7$–$120~\mathrm{M_\odot}$, and $8$–$50~\mathrm{M_\odot}$, corresponding to $k_{\mathrm{CCSN}} = 0.0054$, $0.0089$, and $0.0070~\mathrm{M_\odot^{-1}}$, respectively. The resulting corner plot is presented in Appendix \ref{fig:conmcmc} and the resulting parameters for various $k_{\mathrm{CCSN}}$ values are presented in Appendix Table \ref{tab:mf_theta012_zams_summary}.

\begin{table}[ht]
\centering
\caption{Missing supernova fraction assuming different ZAMS progenitor-mass ranges and the corresponding $k_{\mathrm{CCSN}}$ values. The missing fractions are reported at $z_1=0$, $z_2=2$, and $z_3=4$. Uncertainties correspond to the 16th--84th percentiles (1$\sigma$).}
\label{tab:mf_zams_summary}
\begin{tabular}{ccccc}
\hline
\hline
[$\mathrm{M_\odot}$] &
  $k_{\mathrm{CCSN}}$ &
  $f_{\mathrm{miss}}(z_1)$ [\%] &
  $f_{\mathrm{miss}}(z_2)$ [\%] &
  $f_{\mathrm{miss}}(z_3)$ [\%] \\
  \hline
  $7$--$20$ & 0.0069 & $18^{+8}_{-8}$ & $58^{+9}_{-9}$ & $33^{+23}_{-23}$ \\
  $7$--$50$ & 0.0085 & $34^{+6}_{-6}$ & $66^{+7}_{-7}$ & $46^{+18}_{-18}$ \\
  $7$--$120$ & 0.0089 & $37^{+6}_{-6}$ & $68^{+7}_{-7}$ & $48^{+17}_{-18}$ \\
  \hline
  $8$--$20$ & 0.0054 & $-4^{+10}_{-10}$ & $47^{+11}_{-11}$ & $15^{+29}_{-29}$ \\
  $8$--$50$ & 0.0070 & $20^{+8}_{-8}$ & $59^{+8}_{-8}$ & $34^{+22}_{-22}$ \\
  $8$--$120$ & 0.0074 & $24^{+7}_{-7}$ & $61^{+8}_{-8}$ & $38^{+21}_{-21}$ \\
  \hline
  $9$--$20$ & 0.0043 & $-31^{+12}_{-12}$ & $33^{+13}_{-14}$ & $-6^{+36}_{-36}$ \\
  $9$--$50$ & 0.0059 & $4^{+9}_{-9}$ & $51^{+10}_{-10}$ & $22^{+27}_{-27}$ \\
  $9$--$120$ & 0.0063 & $11^{+9}_{-9}$ & $54^{+9}_{-9}$ & $27^{+25}_{-25}$ \\
\hline
\end{tabular}
\end{table}

A key result of this work is that the newly observed JADES CCSN rates are broadly consistent with the UV+IR SFRD of \citet{Madau2014} for plausible values of $k_{\mathrm{CCSN}}$, requiring no additional missing supernova fraction correction, but fall below expectations when adopting the higher ALMA-based SFRDs reported by \citet{Fujimoto24}. Assuming that \citet{Fujimoto24} SFRDs more accurately capture the total SFRD, this discrepancy can be interpreted as evidence for a substantial population of dust obscured CCSNe missed by the surveys.

The missing CCSN fractions inferred in this work (Table~\ref{tab:mf_zams_summary}) are subject to large uncertainties, but are nevertheless broadly consistent, within errors, with previous estimates based on local dust obscured galaxy populations, such as those presented by \citet{Mattila2012a}. Mattila et al.\ derived a redshift dependent missing fraction that increases from $\sim$19\% at $z=0$ to $\sim$36\% at $z\sim1$, primarily driven by the rising contribution of LIRGs and ULIRGs to the co-moving SFRD (e.g., \citealt{magnelli2011}).

For progenitor mass ranges yielding $k_{\mathrm{CCSN}} \simeq 0.007$--$0.009$, our inferred missing fractions at $z=0$ lie in the range $\sim$20--37\%, consistent with the \cite{Mattila2012a} value of $\sim$19\% within the uncertainties. At higher redshift, our results indicate that the missing supernova fraction rises to a peak of $\sim$59--68\% near $z\approx2$, and then declines to $\sim$34--48\% by $z\approx4$, depending on the assumed progenitor-mass limits (although the $z\approx4$ values carry large uncertainties). While \citet{Mattila2012a} did not extend their analysis beyond $z\approx2$, their framework predicts a substantial contribution to the missing fraction from LIRGs and ULIRGs at $z\gtrsim1$, consistent with the rising trend seen here up to cosmic noon. Although the estimates of \cite{Mattila2012a} were based on an investigation of the CCSN activity in a nearby LIRG, whereas our approach infers the missing supernova fractions by comparing observed volumetric CCSN rates to galaxy luminosity based SFRD measurements, both methodologies indicate that a significant fraction of CCSNe must remain missed by rest-frame optical and near-IR surveys, reaching its maximum around $z\approx2$.

For narrower progenitor-mass ranges with an upper limit of $\sim$$20\,\mathrm{M_\odot}$, corresponding to lower efficiencies $k_{\mathrm{CCSN}} \simeq 0.004$--$0.0055$ (the $8$--$20$ and $9$--$20\,\mathrm{M_\odot}$ ranges), the inferred missing fractions are systematically smaller and, at low redshift, become consistent with zero or even formally negative: at $z=0$ we obtain $-4\pm10$\% for the $8$--$20\,\mathrm{M_\odot}$ range and $-31\pm12$\% for the $9$--$20\,\mathrm{M_\odot}$ range. A negative missing fraction is unphysical and indicates that, for such narrow mass ranges, the observed CCSN rates already meet or exceed the \citet{Fujimoto24} SFRD expectation, leaving no room for a missed population. This suggests that these narrow progenitor ranges are disfavored if a non-negative missing fraction is required at low redshift, in particular the $9$--$20\,\mathrm{M_\odot}$ range, for which the missing fraction is negative at the $\sim$2.5$\sigma$ level. At higher redshift the tension relaxes, with the missing fraction rising to $\sim$33--47\% near $z\approx2$ before declining to $\sim$$-6$--15\% by $z\approx4$, where the values are consistent with zero within the large uncertainties. Taken together, the narrow-range results reinforce the conclusion that a substantial missed CCSN population is required primarily around cosmic noon, while at $z=0$ it is recovered only for the wider progenitor-mass ranges ($k_{\mathrm{CCSN}} \gtrsim 0.007$).

These trends are qualitatively consistent with independent constraints on
dust-obscured star formation. IR and sub-millimeter studies show that
obscured star formation dominates near the cosmic noon and remains significant at higher redshifts \citep[e.g.][]{Zavala2021a, Traina2024, Fujimoto24, Sun2025}. At $z \gtrsim 1$, an increasing fraction of star formation occurs in LIRGs and ULIRGs, containing very large amounts of dust that is expected to fully obscure a substantial fraction of CCSNe in rest-frame optical and even near-infrared searches \citep[e.g.][]{Mannucci2007, Mattila2012a, miluzio2013, Fox2021, mantynen2025}.

An additional systematic uncertainty in the CCSN rate measurements is related to the assumptions about the intrinsic SN luminosity function and its possible evolution with redshift. Intrinsic supernova faintness and dust obscuration are physically distinct effects, but from the perspective of supernova rate calculations both reduce the fraction of CCSNe detectable by a given survey. If the true luminosity function contains a larger fraction of intrinsically faint events than assumed, also an increasing number of CCSNe will remain undetected, leading to an underestimated observed CCSN rate, an effect that is observationally equivalent to using a larger missing supernova fraction.

Observational evidence for a substantial incompleteness is provided by adaptive optics assisted near-infrared searches for supernovae in nearby LIRGs \citep[e.g.][]{mattila2007, Kankare2008, Kool2018}. These efforts indicate that $51$--$75\%$ of CCSNe in local LIRGs are missed even in $K$-band searches, increasing to $85$--$92\%$ for typical optical surveys \citep{mantynen2025}. While these measurements apply specifically to local LIRGs, they illustrate the level of incompleteness that can arise in environments where a significant fraction of dust obscured star formation occurs, motivating the use of missing supernova fraction corrections for any CCSN rate--SFRD comparisons.

\section{Systematic effects and uncertainties}
\label{sec:systematics}
In this section, we discuss the JTS selection effects and effects of the Hubble constant on the CCSN rate - SFRD relation. Additional possible effects that affect the relation as a function of redshift include direct collapse/fallback of the stellar core to a black hole, and the effects of close binary systems and metallicity to stellar evolution. We will explore these effects in our future paper (Vassallo et al. in prep.).

\subsection{JTS selection effects}
\label{sec:systematics_gold}

One of the largest systematic uncertainties in the JTS CCSN rates likely comes from how the supernovae are classified. In the full JTS sample, many events are classified from a single epoch of photometry, which is less reliable than using a spectrum or a multi-epoch light curve and can allow misclassified sources. To test how much this matters, we compare with the JADES ``gold'' sample, whose events are more securely classified using spectra and/or multi-epoch photometry. However, the gold sample is not automatically the better choice, because higher-redshift targets were preferentially followed up, it is biased towards high-$z$ events, whereas the full sample is not. Therefore, the analyzes presented in sections~\ref{sec:LinkingSFR} -- \ref{sec:MF} adopt the full JTS CCSN sample (DeCoursey et al.\ submitted).

 The JTS higher-purity ``gold'' sample volumetric CCSN rates (in units of
$10^{-4}\,h_{70}^{3}\,\mathrm{yr}^{-1}\,\mathrm{Mpc}^{-3}$) are
$3.6^{+2.8}_{-1.7}\,(^{+0.2}_{-0.5})$ ($0.99 \le z < 1.61$),
$8.1^{+5.0}_{-3.2}\,(^{+0.5}_{-0.5})$ ($1.61 \le z < 2.05$),
$6.8^{+3.8}_{-2.5}\,(^{+0.4}_{-0.2})$ ($2.05 \le z < 2.83$), and
$5.2^{+3.1}_{-2.0}\,(^{+1.1}_{-0.4})$ ($2.83 \le z \le 4.45$), with the systematic uncertainties given in parentheses. Because the gold sample contains fewer events per bin, its volumetric rates carry larger statistical uncertainties. Here we repeat the three analyses using the gold JADES rates in place of the full sample, keeping the $z<1$ literature compilation unchanged, to assess whether our conclusions depend on the JTS sample selection.

For the best-fit CCSN efficiency (Section~\ref{sec:LinkingSFR3}), the gold sample
yields results consistent with the full sample within their larger
uncertainties. Adopting the \citet{Madau2014} SFRD and a Salpeter IMF with the
lower progenitor-mass limit fixed at $8\,\mathrm{M_\odot}$, the JADES-only fit
gives $k_{\mathrm{CCSN}} = 0.0049 \pm 0.0013\,\mathrm{M_\odot^{-1}}$ and the
combined JADES$+$literature fit gives
$k_{\mathrm{CCSN}} = 0.0054 \pm 0.0004\,\mathrm{M_\odot^{-1}}$, in close
agreement with one another. The corresponding missing-supernova-corrected values
are $0.0079 \pm 0.0021$ (JADES only) and
$0.0073 \pm 0.0006\,\mathrm{M_\odot^{-1}}$ (combined). For the gold sample,
the \citet{Fujimoto24} SFRD yields systematically lower efficiencies
($k_{\mathrm{CCSN}} \approx 0.0032$--$0.0043\,\mathrm{M_\odot^{-1}}$ without the
missing-supernova correction), while the \citet{Harikane22} SFRD gives
intermediate values ($k_{\mathrm{CCSN}} \approx 0.0041$--$0.0046\,
\mathrm{M_\odot^{-1}}$).

When constructing the cosmic star-formation history from the gold sample rates
(Section~\ref{sec:constrains}), we obtain a median peak redshift of $z_{\rm peak} \approx 3.1$ ($-1.1$ at $1\sigma$),
consistent with the full-sample value
($z_{\rm peak}=2.4^{+1.0}_{-0.6}$). The peak is, however, poorly constrained towards high redshift, because
for $\sim$30\% of the posterior samples the gold-sample SFRD does not turn over below $z=8$, owing to the smaller sample size and flatter shape.
The posterior-predictive tests preserve the same qualitative ranking as the full
sample but at reduced significance, reflecting the weaker constraining power (larger statistical uncertainties) of
the gold rates. Assuming $k_{\mathrm{CCSN}} = 0.0070\,\mathrm{M_\odot^{-1}}$, the
\citet{Madau2014} SFRD is consistent with the CCSN-derived posterior (rejected
only at the 68\% level by the RMS, KS and $\chi^2$ statistics, with $\chi^2$ $p=0.12$), the
\citet{Harikane22} SFRD shows intermediate agreement ($\chi^2$ $p=0.061$), and
the \citet{Fujimoto24} SFRD remains the most disfavored, rejected at the 68\%
and 95\% levels by the $\chi^2$ test ($p=0.032$) and at the 68\% level by the
RMS statistic ($p=0.12$). This is a weaker rejection than for the full sample,
where the \citet{Fujimoto24} SFRD was rejected at up to the 99\% level.

For the missing-fraction analysis (Section~\ref{sec:MF}), the gold sample
reproduces the same redshift trend, rising to a maximum near $z\approx2$ before
declining toward $z\approx4$. Adopting the canonical
$k_{\mathrm{CCSN}} = 0.0070\,\mathrm{M_\odot^{-1}}$ ($8$--$50\,\mathrm{M_\odot}$),
the inferred missing fraction increases from
$f_{\mathrm{miss}}(z{=}0)=20^{+8}_{-8}\%$ to
$f_{\mathrm{miss}}(z{=}2)=55^{+13}_{-13}\%$ and then declines to
$f_{\mathrm{miss}}(z{=}4)=25^{+40}_{-42}\%$. For the lower
$k_{\mathrm{CCSN}} = 0.0054\,\mathrm{M_\odot^{-1}}$ ($8$--$20\,\mathrm{M_\odot}$),
the corresponding values are
$f_{\mathrm{miss}}(z{=}0)=-4^{+10}_{-10}\%$,
$f_{\mathrm{miss}}(z{=}2)=41^{+17}_{-17}\%$, and
$f_{\mathrm{miss}}(z{=}4)=3^{+52}_{-53}\%$. For progenitor ranges
yielding $k_{\mathrm{CCSN}}\simeq 0.007$--$0.009$, the gold-sample missing
fractions span $\sim$20--37\% at $z=0$, $\sim$55--64\% near the $z\approx2$
peak, and $\sim$25--41\% at $z\approx4$.

Overall, the gold and full JADES samples yield consistent results across all
three analyses. The best-fit efficiencies agree well within their $1\sigma$
uncertainties (e.g.\ for the \citet{Madau2014} JADES-only fit,
$0.0049 \pm 0.0013$ for gold versus $0.0049 \pm 0.0008\,\mathrm{M_\odot^{-1}}$
for the full sample), the CCSN-derived SFRD peaks near cosmic noon in both cases,
and the inferred missing fractions agree within their uncertainties at every
redshift. There are no significant differences: the gold rates carry larger uncertainties, which weakens the statistical test significance of higher \citet{Fujimoto24} SFRD with uncorrected CCSN rates (chi squared statistic rejected at up to the 95\% level for the gold sample, compared with the 99\% level for the full sample) and yields a slightly lower missing-fraction peak ($\sim$55--64\% near $z\approx2$,
versus $\sim$59--68\% for the full sample). Adopting the gold sample does not change our conclusions, it mainly reduces the statistical significance of the comparisons, as a consequence of the larger statistical uncertainties of this smaller sample.

\subsection{Effects of Changing the Hubble Constant}
\label{sec:systematics3}
In some studies, Equation \ref{eq:ccsn_sfr_relation} includes an additional factor $h^2$ that multiplies the SFRD (see \citealt{Dahlen2012a} \& \citealt{Strolger2015a}). According to \cite{croton2013}, $h$ is defined as:

\begin{equation}
H_0 = 100 \, h \, \text{km} \, \text{s}^{-1} \, \text{Mpc}^{-1}
\end{equation}

and $h_{70}$ with:

\begin{equation}
h_{70} \equiv \frac{H_0}{70 \text{km}\,\text{s}^{-1}\,\text{Mpc}^{-1}} = 1 \qquad 
\end{equation}
assuming $H_0 = 70\, \text{km}\,\text{s}^{-1}\,\text{Mpc}^{-1}$.
We do not include $h$ in Equation \ref{eq:ccsn_sfr_relation} and instead describe how a change in $h$ should be treated based on the guidelines of \cite{croton2013}. The inferred comoving volume scales as $h^{-3}$, and therefore volumetric CCSN rates (per comoving Mpc$^3$) scale as $h^{3}$. The SFRD inferred from luminosity densities has a dependence on $h^{-2}$ from luminosity and on $h^{-3}$ from volume, and thus scales as $h$. When comparing CCSN rates or SFRDs derived using methods with different $h$-dependences, such as theoretical and observational approaches, it is very important not to transform one to match the $h$-scaling of the other. Instead, a consistent $h$-value should be selected, and both observables should be scaled to that independently, and only then compared with each other. These guidelines are described in detail in \cite{croton2013}. In our study the observed and expected CCSN rates determined from the cosmic SFH have been estimated in or transformed to the same cosmology (unless specified otherwise).

If we change the value of $h$, the derived value of $k$ also changes accordingly. This is because $k = \mathrm{CCSNR} / \mathrm{SFRD}$ scales as $h^2$, given that CCSN rates scale as $h^3$ and SFRD as $h$. To quantify the importance of this effect, we evaluated the change in $k$ when changing from $h = 0.70$ to $h = 0.65$ or $h = 0.75$. These changes result in $-14\%$ and $+15\%$ effects for $k$, respectively, due to the $\left(h/h_{70}\right)^2$ scaling. Changing the value of $\Omega_m$ and $\Omega_{\lambda}$ slightly affect the SN rates as well, but is not considered here.

\section{Conclusions}
\label{sec:conclusions}
We have investigated the connection between CCSN
rates and SFRDs from the local Universe
to $z \sim 5$, combining the new highest redshift results for volumetric CCSN rates of the JADES Transient Survey with a compilation of previous volumetric CCSN rates. By comparing CCSN rates with the SFRDs inferred from galaxy luminosities, and by inverting the
CCSN--SFRD relation, we have assessed both the consistency of CCSN rates as
tracers of cosmic star formation and the implications of recent
ALMA-based millimeter SFRD measurements for high-redshift CCSN rates.

Our main findings can be summarized as follows:

\begin{enumerate}
  \item When adopting the dust extinction-corrected UV+IR SFRD of
  \citet{Madau2014} and the Salpeter IMF, the observed CCSN rates are broadly
  consistent with expectations across the full redshift range probed. 
  Fitting the combined JADES and literature CCSN rates yields a CCSN production
  efficiency, $k_{\mathrm{CCSN}} \simeq 0.0053\,\mathrm{M_\odot^{-1}}$,
  consistent with an upper ZAMS progenitor-mass of $\sim$ $19\,\mathrm{M_\odot}$ if assuming a lower ZAMS progenitor mass of $8\,\mathrm{M_\odot}$. Applying the missing-supernova correction increases the inferred efficiency. For the combined CCSN sample it yields $k_{\mathrm{CCSN}} = 0.0073\,\mathrm{M_\odot^{-1}}$, corresponding to an upper ZAMS progenitor-mass of $\sim$ $86\,\mathrm{M_\odot}$ ($\gtrsim 40\,\mathrm{M_\odot}$ within $1\sigma$), and for the corrected JADES-only rates an even higher efficiency of $k_{\mathrm{CCSN}} = 0.0078\, \mathrm{M_\odot^{-1}}$ is obtained, which is consistent with a wide progenitor mass range of $\sim 8$--$120\,\mathrm{M_\odot}$. Both corrected values thus point to wide progenitor mass ranges covering most of the massive-star regime, leaving little space for stellar cores collapsing to a black hole without a bright supernova, whereas the uncorrected value implies a narrow range. Our results therefore illustrate the importance of correcting both the CCSN rates and SFRDs consistently for effects of dust obscuration.
  
  \item By inverting the CCSN--SFRD relation at fixed $k_{\mathrm{CCSN}}$, we reconstruct the cosmic star-formation history directly from the observed CCSN rates uncorrected for missing SNe. The CCSN-derived SFRD peaks at $z_{\rm peak} = 2.4^{+1.0}_{-0.6}$, or $z_{\rm peak} = 2.3^{+0.6}_{-0.5}$ when excluding the $\sim$10\% of posterior samples that peak beyond the redshift range covered by the data ($z>5$), independent of adopted $k_{\mathrm{CCSN}}$, both in agreement with galaxy luminosity-based determinations such as \citet{Madau2014} ($z \sim 1.9$). Posterior-predictive tests of the literature SFRDs against our CCSN-derived posterior, performed for both $k_{\mathrm{CCSN}} = 0.0070$ and $0.0054,\mathrm{M_\odot^{-1}}$, show that the \citet{Madau2014} SFRD is consistent with the observed CCSN rates (rejected only at the 68\% level), the \citet{Harikane22} SFRD shows slightly less agreement, and the $1.6\times$ higher \citet{Fujimoto24} SFRD is the most disfavored, rejected by the $\chi^2$ statistic at up to 99\% confidence for $k_{\mathrm{CCSN}} = 0.0070,\mathrm{M_\odot^{-1}}$ (up to 95\% for $0.0054,\mathrm{M_\odot^{-1}}$). Since the shape-sensitive KS test shows at most marginal tension for any model, this disagreement reflects the overall normalization of the CCSN rates rather than their evolution over the redshift.

  \item If the higher cosmic SFRDs inferred from millimeter ALMA observations
  \citep[e.g.,][]{Fujimoto24} are adopted, the observed CCSN rates imply a
  substantial population of dust-obscured CCSNe missed by current surveys including the JTS. The
  inferred missing SN fraction depends on the assumed CCSN production efficiency,
  $k_{\mathrm{CCSN}}$. For example, adopting the commonly used value
  $k_{\mathrm{CCSN}} = 0.0070\,\mathrm{M_\odot^{-1}}$ ($\sim
  8$--$50\,\mathrm{M_\odot}$ for a Salpeter IMF) yields an inferred missing
  CCSN fraction that increases from
  $f_{\mathrm{miss}}(z{=}0)=20^{+8}_{-8}\%$ to a peak of
  $f_{\mathrm{miss}}(z{=}2)=59^{+8}_{-8}\%$, and then declines to
  $f_{\mathrm{miss}}(z{=}4)=34^{+22}_{-22}\%$.
  The required missing SN-correction is reduced if a lower CCSN production
 efficiency is adopted. Using $k_{\mathrm{CCSN}} =
 0.0054\,\mathrm{M_\odot}^{-1}$ ($\sim 8$--$20\,\mathrm{M_\odot}$ for a
 Salpeter IMF) a value similar to the best-fit efficiency obtained for the
 combined JADES+literature CCSN rate sample gives
 $f_{\mathrm{miss}}(z{=}0)=-4^{+10}_{-10}\%$ (consistent with zero), increasing to
 $f_{\mathrm{miss}}(z{=}2)=47^{+11}_{-11}\%$ and then declining to
 $f_{\mathrm{miss}}(z{=}4)=15^{+29}_{-29}\%$.

\item According to our analysis, constraining the IMF from CCSN rates alone is very challenging, and other information, such as relative rates between different CCSN subtypes, would be required. This is because varying the IMF alters both the CCSN production efficiency ($k_{\mathrm{CCSN}}$) and the galaxy luminosity-based SFRDs, and these two effects cancel out largely. Switching from a Salpeter IMF to an extreme top-heavy IMF such as the one discussed in \cite{harikane23} for z $\sim \,10 $--$ 16$ would only change the expected CCSN rates by approximately $\sim 35\%$. Such an effect would easily remain undetected given the large uncertainties in the current CCSN rates estimated at the highest redshifts and other uncertainties including progenitor mass ranges and dust obscuration. In addition, in the redshift range considered in this work ($z = 0$--5), such extreme IMFs are unlikely to be relevant. For the top heavy IMF of \cite{Davee2008} proposed for $z=2$ the combined effect on the CCSN - SFRD relation remains below 5\%.

\end{enumerate}

We find that reasonable variations in the assumptions of the IMF and the cosmological parameters do not significantly alter these conclusions. Although these choices affect the efficiency of the CCSN--SFRD relation, they alone cannot reconcile the observed CCSN rates with the highest inferred SFRD densities at high redshift. It is also important to note that the missing CCSN fraction required to match the ALMA-based millimeter SFRDs is degenerate with assumptions about the CCSN progenitor mass range and the corresponding CCSN--SFRD conversion factor, $k_{\mathrm{CCSN}}$. Adopting a higher low-mass cutoff for CCSN progenitors reduces the expected CCSN rate per unit SFRD, and therefore lowers the missing fraction required to reconcile CCSN rates with a given SFRD, while more inclusive progenitor mass ranges increase the missing fraction required. Although this degeneracy affects $f_{\rm miss}(z)$, it does not alter the qualitative result that a substantial population of CCSNe is missed at high redshifts.

Overall, our results indicate that CCSN rates provide an independent probe of the cosmic star-formation history, but that their interpretation depends critically on the completeness of both CCSN samples and galaxy luminosity-based SFRD measurements. Future wide-area time-domain surveys with improved infrared sensitivities, together with continued JWST observations, will be essential to obtain more accurate CCSN rates and extending them to higher redshifts, improving constraints on systematics in the CCSN rate–SFR relation throughout the history of galaxy evolution.

\begin{acknowledgments}
This work is based on observations made with the NASA/ESA/CSA James Webb Space Telescope. The data were obtained from the Mikulski Archive for Space Telescopes at the Space Telescope Science Institute, which is operated by the Association of Universities for Research in Astronomy, Inc., under NASA contract NAS 5-03127 for JWST. These observations are associated with program \#1180 and 6541. The specific JWST observations analyzed can be accessed via \dataset[DOI: 10.17909/c4qk-xv53]{https://doi.org/10.17909/c4qk-xv53}. The STScI TSST group acknowledges partial support from JWST-GO-06541, JWST-GO-06585, and JWST-GO-05324.

We thank the anonymous referee whose comments and
suggestions led to significant improvements of the paper. CV acknowledges helpful discussions with Mathieu Renzo, Danny Steeghs, and Elizabeth Stanway. CV, SM and TMR acknowledge financial support from the Research Council of Finland project 350458. CV acknowledges financial support from the Vilho, Yrjö ja Kalle Väisälä foundation. AJB acknowledges funding from the FirstGalaxies Advanced Grant from the European Research Council (ERC) under the European Union’s Horizon 2020 research and innovation programme (Grant agreement No.~789056). IM \& EK acknowledge financial support from the Emil Aaltonen foundation. AJC gratefully acknowledges support from the Cosmic Dawn Center through the DAWN Fellowship. The Cosmic Dawn Center (DAWN) is funded by the Danish National Research Foundation under grant No.~140. DJE is supported as a Simons Investigator. EE, ZJ, BDJ, BER, DJE, and CNAW acknowledge support from the JWST/NIRCam contract to the University of Arizona (NAS 5-02105). BDJ and DJE acknowledge support from JWST Program~3215. Support for Program~3215 was provided by NASA through a grant from the Space Telescope Science Institute, which is operated by the Association of Universities for Research in Astronomy, Inc., under NASA contract NAS 5-03127. RH acknowledges that funding for this research was provided by the Johns Hopkins University Institute for Data Intensive Engineering and Science (IDIES). BER acknowledges use of the lux supercomputer at UC Santa Cruz, funded by NSF MRI grant AST~1828315. RM acknowledges support by the Science and Technology Facilities Council (STFC), by the ERC through Advanced Grant~695671 (QUENCH), and by the UKRI Frontier Research grant RISEandFALL. RM also acknowledges funding from a research professorship from the Royal Society. ST acknowledges support by the Royal Society Research Grant G125142. CCW acknowledges that the research is supported by NOIRLab, which is managed by the Association of Universities for Research in Astronomy (AURA) under a cooperative agreement with the National Science Foundation.

\end{acknowledgments}

\begin{figure*}
    \centering
    \includegraphics[width=0.480\textwidth]{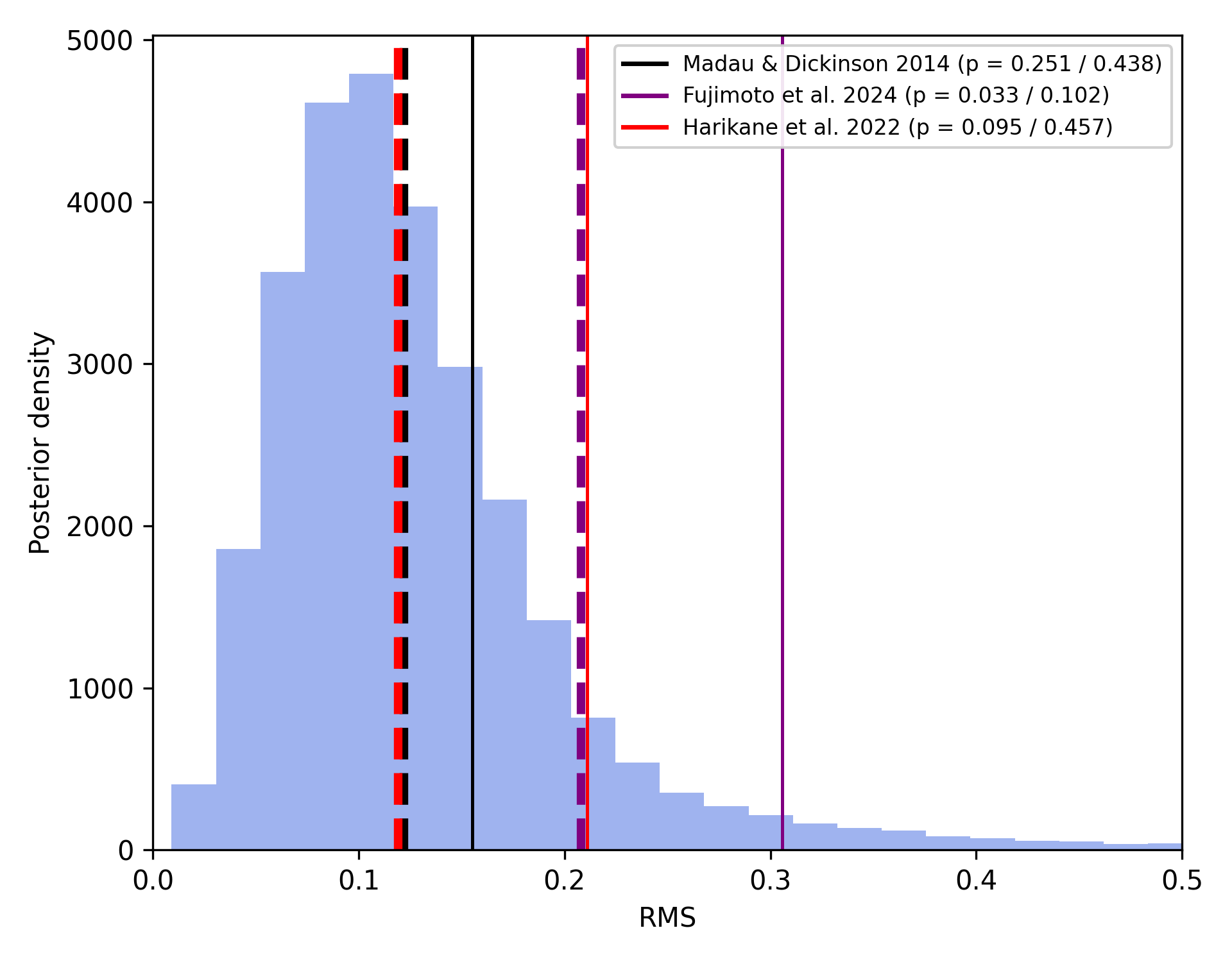}\hfill
    \includegraphics[width=0.48\textwidth]{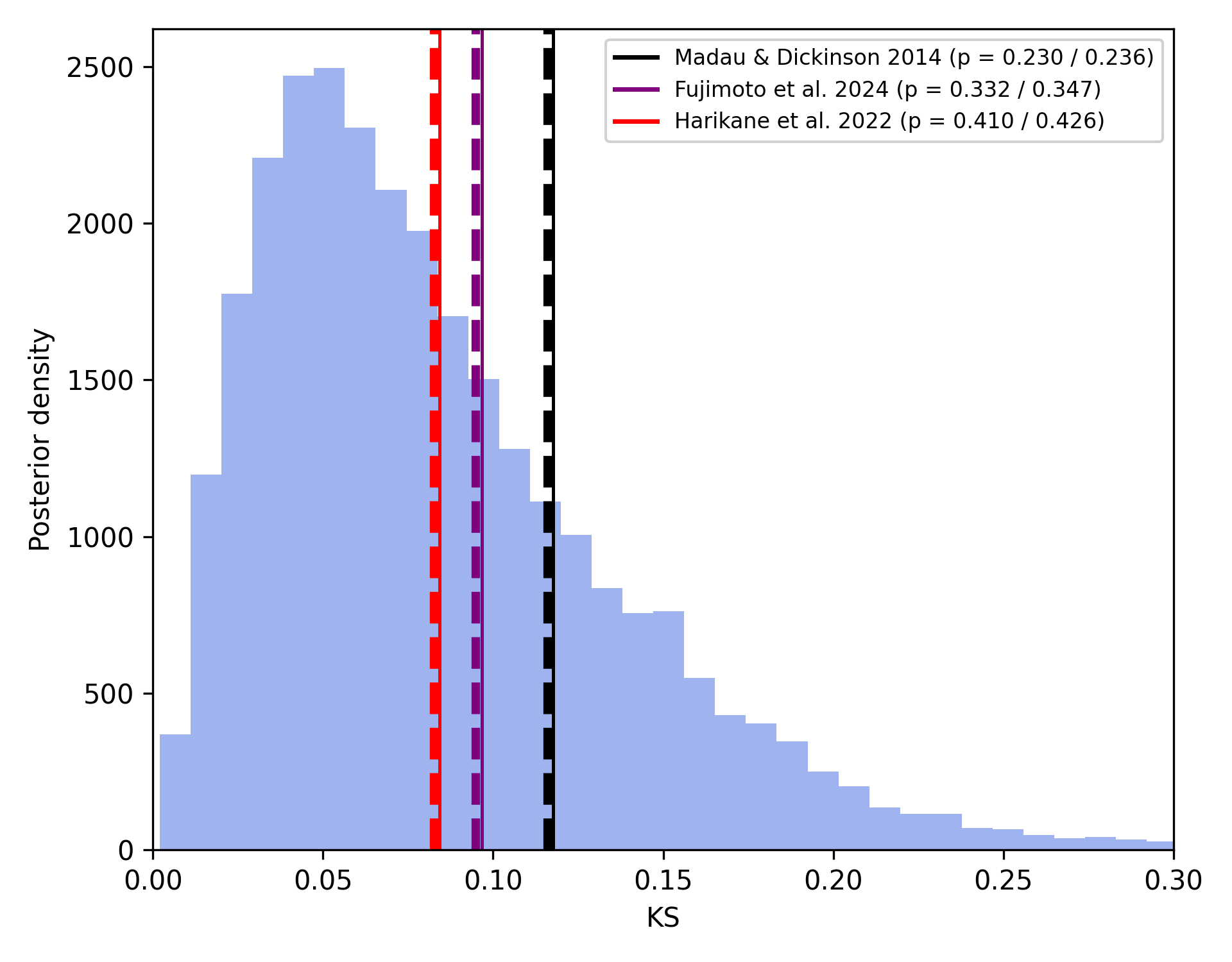}

    \vspace{0.5cm}

    \includegraphics[width=0.48\textwidth]{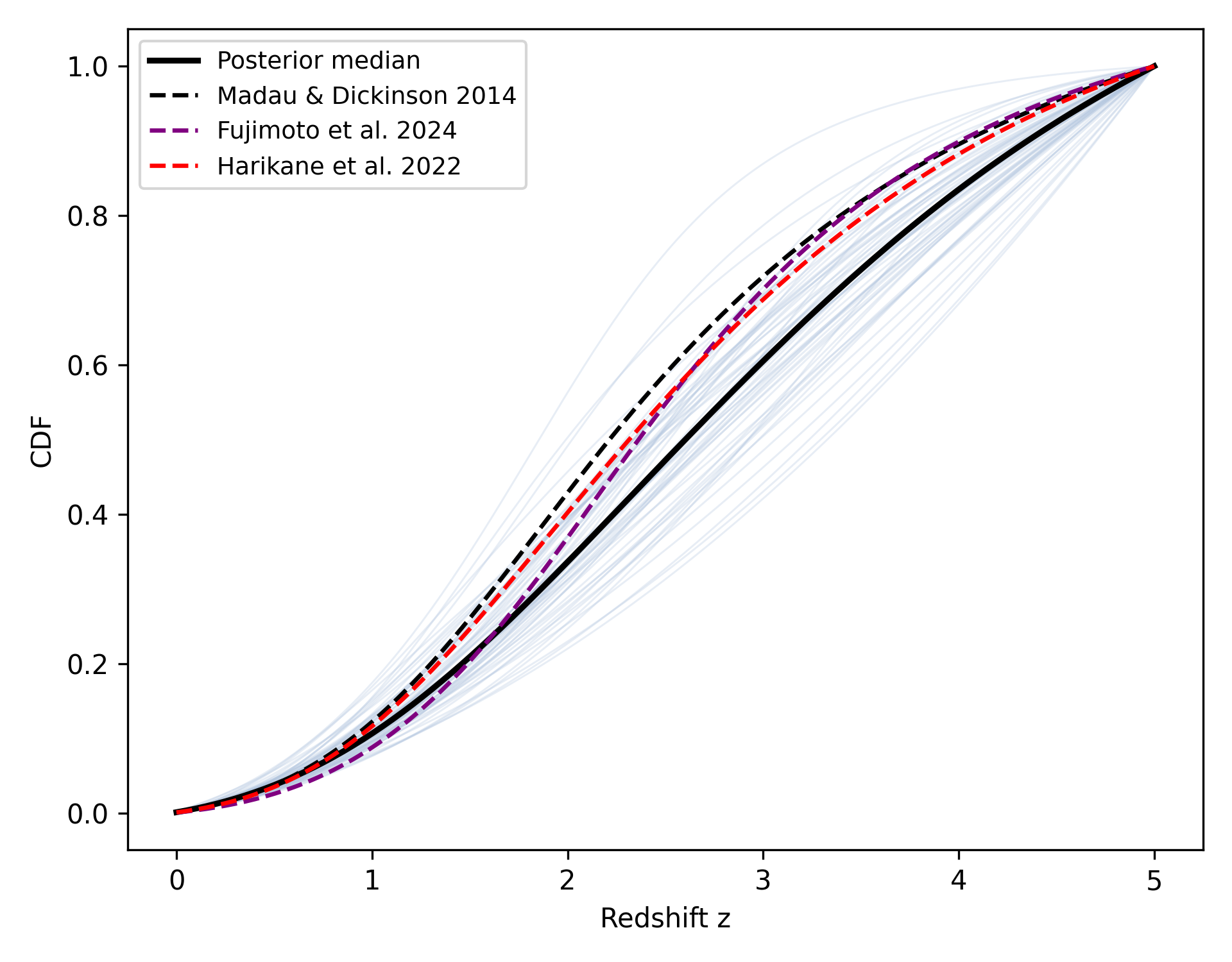}\hfill
    \includegraphics[width=0.48\textwidth]{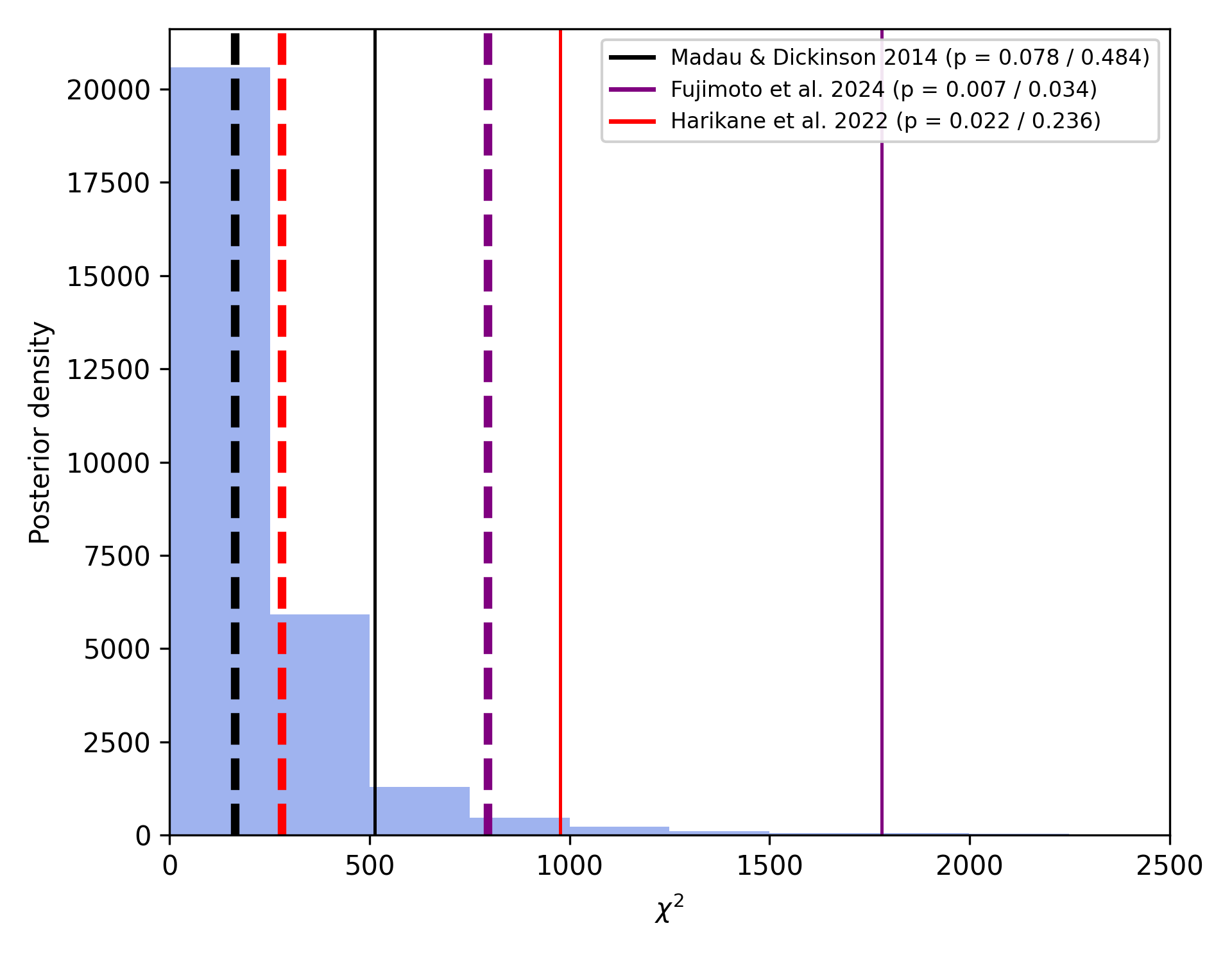}

    \caption{
    Statistical comparison between the posterior CCSN rate distribution and literature SFRD models.
    Top left: RMS test.
    Top right: KS shape test.
    Bottom left: Corresponding cumulative distribution functions used in the KS test.
    Bottom right: Posterior predictive $\chi^2$ test using the posterior spread as the uncertainty.
    The literature models are \citet{Madau2014} (black), \citet{Fujimoto24} (purple),
    and \citet{Harikane22} (red). In the RMS, KS, and $\chi^2$ panels each model is shown for
    two conversion factors, $k_{\mathrm{CCSN}} = 0.0070\,M_{\odot}^{-1}$ (solid lines) and
    $k_{\mathrm{CCSN}} = 0.0054\,M_{\odot}^{-1}$ (dashed lines), and the associated $p$-values
    are quoted as ($0.0070$ / $0.0054$).
    }
    \label{fig:stat_tests}
\end{figure*}

\begin{figure}[b]
    \centering
    \includegraphics[width=0.8\textwidth]{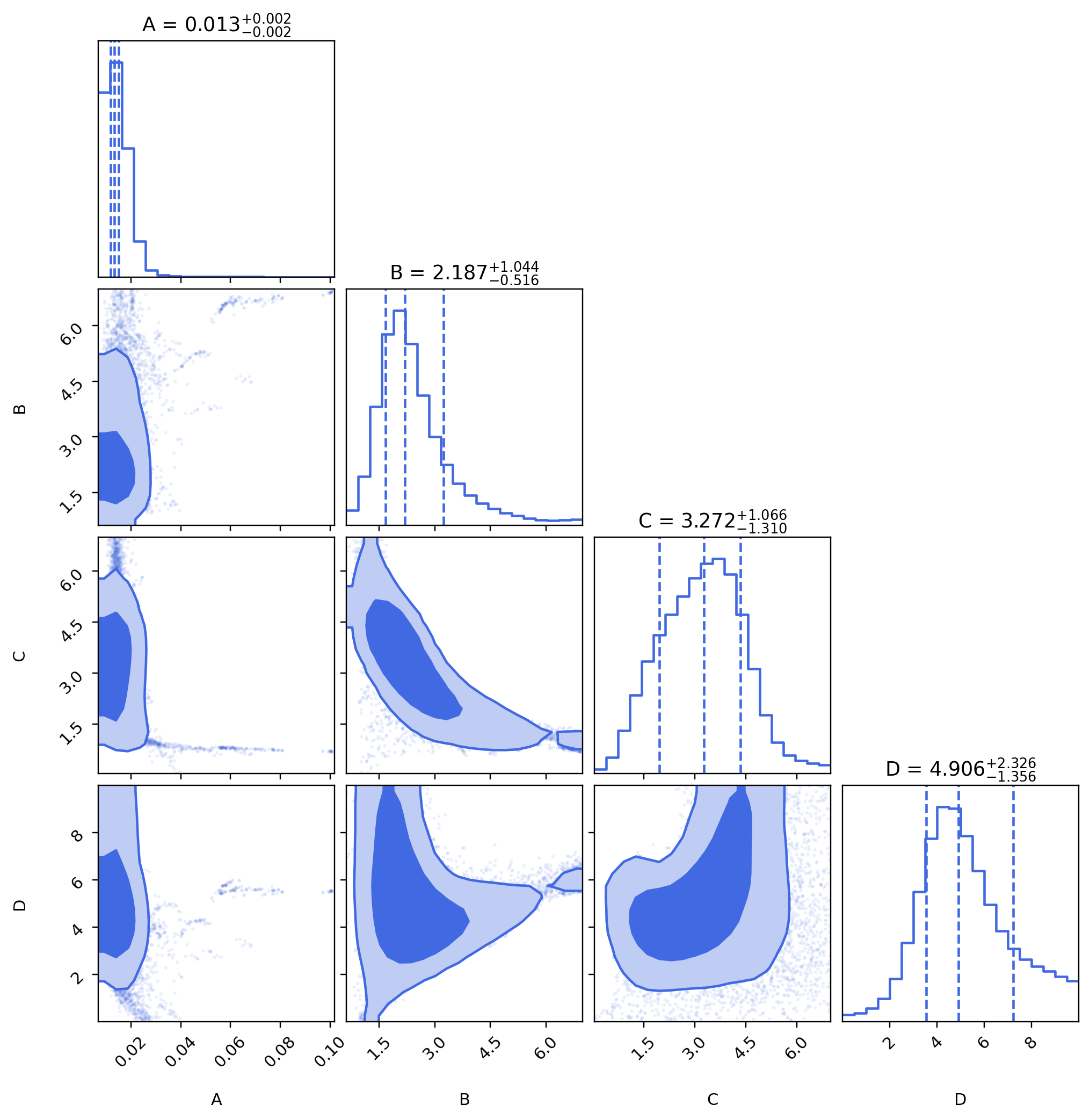}
    \caption{Posterior distributions of the parameters for the SFRD. This uses non-missing supernova correction observed CCSN rates and flat priors for all parameters. This fit uses the conversion factor $k_{\mathrm{CCSN}} = 0.0070 M_{\odot}^{-1}$.}
    \label{fig:bestsfrs}
\end{figure}

\begin{figure}[hbt]
    \centering
    \includegraphics[width=0.49\textwidth]{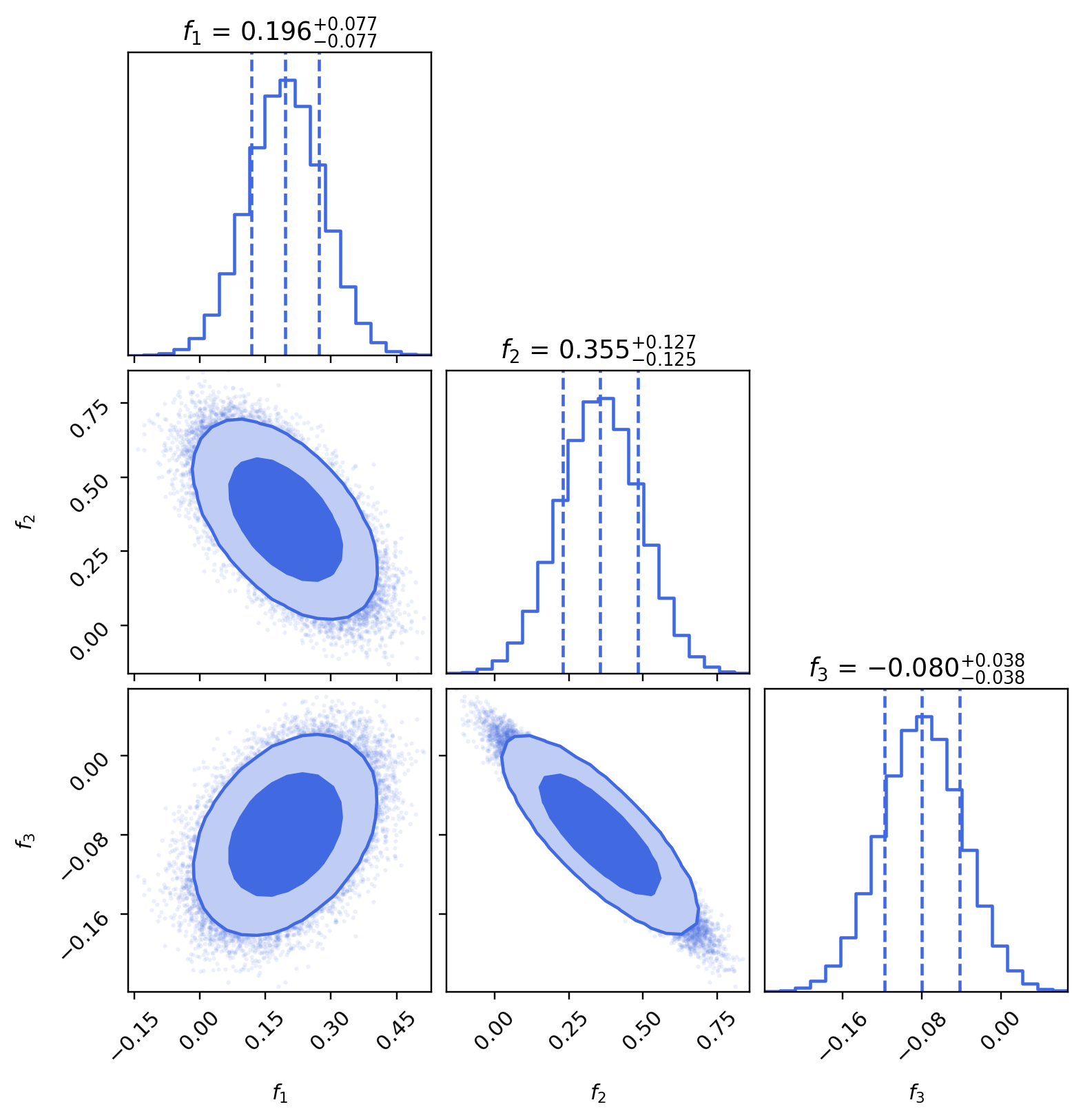}
    \caption{Posterior distributions of the parameters for the continous missing fraction. This uses Fujimoto SFRD converted to Salpeter IMF and $k_{\mathrm{CCSN}} = 0.0070 M_{\odot}^{-1}$.}
    \label{fig:conmcmc}
\end{figure}

\appendix
 Here we present Appendix Tables~\ref{tab:host_extinction}, \ref{tab:csfh_zams_summary}, and \ref{tab:mf_theta012_zams_summary}, together with Appendix Figures~\ref{fig:stat_tests}, \ref{fig:bestsfrs}, and \ref{fig:conmcmc}. Appendix Table~\ref{tab:host_extinction} summarizes how host-galaxy extinction corrections are treated in each CCSN rate survey included in our compilation. Appendix Figure~\ref{fig:bestsfrs} shows the corner plot for the SFRD parameters $A$, $B$, $C$, and $D$ obtained from an MCMC analysis assuming a Salpeter IMF and a fixed CCSN production efficiency $k_{\mathrm{CCSN}}$, under flat priors the posteriors are well constrained over the explored parameter ranges, with the dashed lines marking the posterior medians and 1$\sigma$ (68\%) credible intervals. Appendix Table~\ref{tab:csfh_zams_summary} lists the corresponding CSFH parameter constraints for the different assumed ZAMS progenitor-mass ranges and their fixed $k_{\mathrm{CCSN}}$ values. Appendix Figure~\ref{fig:stat_tests} presents the posterior-predictive statistical tests (the RMS, KS, and $\chi^2$ statistics, together with the corresponding cumulative distribution functions) comparing our CCSN-derived posterior rate distribution with the literature SFRD models of \citet{Madau2014}, \citet{Harikane22}, and \citet{Fujimoto24}, for the two conversion factors $k_{\mathrm{CCSN}} = 0.0070$ and $0.0054,\mathrm{M_\odot^{-1}}$, the corresponding numerical results are summarized in Table~\ref{tab:posterior_tests}. Appendix Figure~\ref{fig:conmcmc} presents the corner plot for the missing-SN-fraction analysis, showing the parameters $f_1$, $f_2$, and $f_3$ derived under the same Salpeter-IMF assumptions and using the $\rho_{\mathrm{SFR}}(z)$ model of \citet{Fujimoto24} transformed to a Salpeter IMF, adopting flat priors, these parameters are likewise well constrained within their allowed ranges, with dashed lines indicating the posterior medians and 1$\sigma$ credible intervals. Appendix Table~\ref{tab:mf_theta012_zams_summary} lists the inferred missing-fraction parameters for the set of ZAMS progenitor-mass ranges and associated fixed $k_{\mathrm{CCSN}}$ values.

\begin{table*}
\centering
\caption{Treatment of host-galaxy extinction in published CCSN rate measurements.}
\label{tab:host_extinction}
\begin{tabular}{l c l}
\hline
Reference & Host ext. corr.? & Notes \\
\hline

C. DeCoursey submitted & Yes &
\notecell{Host galaxy extinction modeled using assumed $A_V$ distribution (\cite{kelly2012}) for CCSNe when computing control times and detection efficiencies. No separate correction for heavily obscured CCSNe in LIRGs/ULIRGs.} \\

\cite{Strolger2015a} & Yes &
\notecell{Host galaxy extinction modeled using assumed $A_V$ distribution (\cite{hatano}) for CCSNe when computing control times and detection efficiencies, with an additional correction applied for CCSNe missed (\cite{Mattila2012a}) in dusty star forming galaxies (including LIRGs/ULIRGs), which are incorporated into the systematic uncertainties.} \\

\cite{Petrushevska2016} & Yes &
\notecell{Host galaxy extinction correction applied using Monte Carlo simulations with a fixed extinction distribution (\cite{rodney}), no separate correction for heavily obscured CCSNe in LIRGs/ULIRGs.} \\

\cite{Dahlen2012a} & Yes &
\notecell{Three CCSN rates reported: (i) uncorrected, (ii) corrected for host-galaxy extinction using Monte Carlo extinction models in normal galaxies (\cite{riello}), and (iii) fully corrected including an additional redshift-dependent correction for CCSNe missed (\cite{Mattila2012a}) in dusty star-forming galaxies (LIRGs/ULIRGs).} \\

\cite{Melinder2012} & Yes &
\notecell{Three CCSN rates reported: (i) uncorrected, (ii) corrected for host-galaxy extinction using Monte Carlo extinction models in normal galaxies (\cite{riello}), and (iii) fully corrected including an additional redshift-dependent correction for CCSNe missed (\cite{Mattila2012a}) in dusty star-forming galaxies (LIRGs/ULIRGs).} \\

\cite{Graur2011} & Yes &
\notecell{Host galaxy extinction treated in detectability simulations only (using model from \cite{neill06}), no correction for heavily obscured CCSNe.} \\

\cite{Cappellaro2005} & No &
\notecell{No host galaxy extinction correction applied, CCSN rate is an optical lower limit due to uncorrected dust extinction.} \\

\cite{Botticella2008} & Yes &
\notecell{Statistical host galaxy extinction correction applied via Monte Carlo modeling of SN and dust distributions (\cite{riello}), rates include extinction corrections.} \\

 \cite{cappellaro2015} & Yes &
\notecell{Extinction modeled (using \cite{neill06} model) in SN template fitting and detection efficiency calculations, no correction applied for heavily dust-obscured CCSNe.} \\

\cite{Graur2015} & Yes &
\notecell{Host galaxy extinction included statistically through dust-corrected CCSN luminosity functions and an assumed $A_V$ distribution \cite{rodney} in detection-efficiency simulations, no recovery of heavily obscured CCSNe.} \\

\cite{taylorr} & Yes &
\notecell{Statistical host galaxy extinction correction applied using the \cite{hatano} inclination based $A_V$ model, small correction treated as systematic, with no compensation for heavily obscured CCSNe.} \\

\cite{Frohmaier2021a} & No &
\notecell{Host galaxy extinction not corrected, extinction is included via observed CCSN light-curve templates. No correction for heavily obscured CCSNe.} \\

\cite{perley2020} & No &
\notecell{No explicit host galaxy extinction correction applied, only Galactic foreground extinction is corrected. Events with high Galactic extinction ($A_V > 1$ mag) are excluded.} \\

\cite{Pessi2025} & No &
\notecell{Observed optical CCSN rate only, no correction for host galaxy extinction or correction for heavily obscured events.} \\

\cite{cappellaro1999} & Partial &
\notecell{CCSN rates are not corrected with an explicit $A_V$ distribution per SN. Instead, host extinction enters through the empirical “inclination bias” and “nuclear bias” corrections (see \cite{cappellaro1999}) applied to the detection efficiency and control time.} \\

\cite{li2011} & Partial &
\notecell{Observed optical CCSN rate only, host galaxy extinction partially compensated by using SN luminosity functions also not corrected for host galaxy extinction, no correction for heavily obscured CCSNe.} \\

\cite{Ma2025} & Yes &
\notecell{Statistical host galaxy extinction (based on \cite{holwerda}) correction included via Monte Carlo detectability modeling, residual obscured CCSNe contribute to the systematic uncertainty.} \\

\cite{Botticella2012} & No &
\notecell{Nearby ($<11$ Mpc) sample mitigates dust bias through proximity, so no explicit host-galaxy extinction correction is applied.} \\

\cite{Mattila2012a} & No &
\notecell{Nearby ($<15$ Mpc) sample includes CCSNe with $A_V \sim 4$--5 mag, proximity enables detection of also heavily obscured events, so no explicit host-galaxy extinction correction is applied.}\\
\hline
\end{tabular}
\end{table*}

\begin{table}[ht]
  \centering
  \caption{Posterior medians and 16th--84th percentiles (1$\sigma$) for the CSFH parametrization in Equation \ref{eq:sfr_params}, assuming different ZAMS progenitor-mass ranges and the corresponding fixed $k_{\mathrm{CCSN}}$ values (Salpeter IMF).
  The different $k_{\mathrm{CCSN}}$ values only affect $A$. The negligible differences in the other parameters are due to small randomness in the MCMC method.}
  \label{tab:csfh_zams_summary}
  \begin{tabular}{c c c c c c}
  \hline
  \hline
    [$\mathrm{M_\odot}$] & $k_{\mathrm{CCSN}}$ & $A$ [$\mathrm{M_\odot}\,\mathrm{yr}^{-1}\,\mathrm{Mpc}^{-3}$] & $B$ & $C$ & $D$ \\
    \hline
    $7$--$20$ & 0.0069 & $0.013^{+0.002}_{-0.001}$ & $2.2^{+1.0}_{-0.5}$ & $3.3^{+1.1}_{-1.3}$ & $4.8^{+2.4}_{-1.4}$ \\
    $7$--$50$ & 0.0085 & $0.011^{+0.001}_{-0.001}$ & $2.2^{+1.0}_{-0.5}$ & $3.3^{+1.1}_{-1.3}$ & $4.9^{+2.4}_{-1.4}$ \\
    $7$--$120$ & 0.0089 & $0.010^{+0.001}_{-0.001}$ & $2.2^{+1.0}_{-0.5}$ & $3.2^{+1.1}_{-1.3}$ & $4.8^{+2.3}_{-1.4}$ \\
    \hline
    $8$--$20$ & 0.0054 & $0.017^{+0.002}_{-0.002}$ & $2.2^{+1.0}_{-0.5}$ & $3.3^{+1.1}_{-1.3}$ & $4.9^{+2.4}_{-1.4}$ \\
    $8$--$50$ & 0.0070 & $0.013^{+0.002}_{-0.002}$ & $2.2^{+1.0}_{-0.5}$ & $3.3^{+1.1}_{-1.3}$ & $4.9^{+2.3}_{-1.4}$ \\
    $8$--$120$ & 0.0074 & $0.013^{+0.002}_{-0.001}$ & $2.2^{+1.0}_{-0.5}$ & $3.3^{+1.1}_{-1.3}$ & $4.9^{+2.4}_{-1.4}$ \\
    \hline
    $9$--$20$ & 0.0043 & $0.022^{+0.003}_{-0.002}$ & $2.2^{+1.0}_{-0.5}$ & $3.3^{+1.1}_{-1.3}$ & $4.9^{+2.4}_{-1.4}$ \\
    $9$--$50$ & 0.0059 & $0.016^{+0.002}_{-0.002}$ & $2.2^{+0.9}_{-0.5}$ & $3.3^{+1.0}_{-1.3}$ & $4.9^{+2.4}_{-1.4}$ \\
    $9$--$120$ & 0.0063 & $0.015^{+0.002}_{-0.002}$ & $2.2^{+1.0}_{-0.5}$ & $3.3^{+1.1}_{-1.3}$ & $4.9^{+2.4}_{-1.4}$ \\
  \hline
  \end{tabular}
\end{table}
  
\begin{table}[ht]
\centering
\caption{Posterior medians and 16th--84th percentiles (1$\sigma$) for the 2nd degree missing fraction model (Equation \ref{eq:MF_params}) parameters $f_1$, $f_2$, and $f_3$, assuming different ZAMS progenitor-mass ranges and the corresponding fixed $k_{\mathrm{CCSN}}$ values (Salpeter IMF).}
\label{tab:mf_theta012_zams_summary}
\begin{tabular}{c c c c c}
\hline
\hline
[$\mathrm{M_\odot}$] & $k_{\mathrm{CCSN}}$ & $f_1$ & $f_2$ & $f_3$ \\
\hline
  $7$--$20$ & 0.0069 & $0.18^{+0.08}_{-0.08}$ & $0.36^{+0.13}_{-0.13}$ & $-0.08^{+0.04}_{-0.04}$ \\
  $7$--$50$ & 0.0085 & $0.34^{+0.06}_{-0.06}$ & $0.29^{+0.10}_{-0.10}$ & $-0.07^{+0.03}_{-0.03}$ \\
  $7$--$120$ & 0.0089 & $0.37^{+0.06}_{-0.06}$ & $0.28^{+0.10}_{-0.10}$ & $-0.06^{+0.03}_{-0.03}$ \\
  \hline
  $8$--$20$ & 0.0054 & $-0.04^{+0.10}_{-0.10}$ & $0.46^{+0.16}_{-0.16}$ & $-0.10^{+0.05}_{-0.05}$ \\
  $8$--$50$ & 0.0070 & $0.20^{+0.08}_{-0.08}$ & $0.36^{+0.13}_{-0.13}$ & $-0.08^{+0.04}_{-0.04}$ \\
  $8$--$120$ & 0.0074 & $0.24^{+0.07}_{-0.07}$ & $0.34^{+0.12}_{-0.12}$ & $-0.08^{+0.04}_{-0.04}$ \\
  \hline
  $9$--$20$ & 0.0043 & $-0.31^{+0.12}_{-0.12}$ & $0.57^{+0.20}_{-0.20}$ & $-0.13^{+0.06}_{-0.06}$ \\
  $9$--$50$ & 0.0059 & $0.04^{+0.09}_{-0.09}$ & $0.42^{+0.15}_{-0.15}$ & $-0.10^{+0.05}_{-0.04}$ \\
  $9$--$120$ & 0.0063 & $0.11^{+0.09}_{-0.09}$ & $0.39^{+0.14}_{-0.14}$ & $-0.09^{+0.04}_{-0.04}$ \\
\hline
\end{tabular}
\end{table}
\newpage
\bibliographystyle{aasjournalv7}
\bibliography{SNrategradu}
\end{document}